\documentclass[12pt]{JHEP3}

\RequirePackage{epsf}
\RequirePackage{longtable}

\newcommand{\roughly}[1]{\mathrel{\raise.3ex\hbox{$#1$\kern-0.85em
\lower1ex\hbox{$\sim$}}}}

\newcommand{\lsim}{\roughly<}

\title{Quantum Gravity in Everyday Life:\\
General Relativity as an Effective Field Theory\footnote{To appear
in {\it Living Reviews of Relativity}.}}

\author{C.P.\ Burgess \\
        Physics Department, McGill University \\
        3600 University Street, Montr{\'e}al \\
        Qu{\'e}bec, Canada, H3A 2T8. \\
        e-mail: cliff@physics.mcgill.ca \\
        http://www.physics.mcgill.ca/\protect\~\protect cliff \\
\\
\small{(last modified: 9 October 2003)} }

\date{}

\abstract{This article is meant as a summary and introduction to
the ideas of effective field theory as applied to gravitational
systems, ideas which provide the theoretical foundations for the
modern use of General Relativity as a theory from which precise
predictions are possible.}


\begin{document}



\section{Introduction}
\label{section:introduction}
Quantum mechanics and General Relativity were discovered within a
decade of one another close to a century ago, and almost
immediately the search for a quantum theory of gravity had begun.
Ever since it has been a puzzle as to what theoretical framework
might ultimately reconcile these two theories with one another.
This reconciliation has proven to be difficult to achieve, and
although several promising proposals now exist none can yet claim
complete theoretical and experimental vindication.

\subsection{Against the Split Brain}
The long-standing nature of this difficulty has driven some
physicists to a state of intellectual despair, wherein they
conclude that a crisis exists in physics which might be called the
crisis of the \emph{split brain}. On one hand, quantum mechanics
(and its offspring quantum field theory) provides an incredibly
successful description of all known non-gravitational phenomena,
with agreement between predictions and experiment sometimes taking
place at the part-per-billion level \cite{QED,EWdata}. On the
other hand, classical general relativity is also extremely
successful, with its predictions being well tested within the
solar system and for some binary pulsar systems \cite{GRdata}.
(The cosmological evidence for dark matter and dark energy is
sometimes proposed as indicating the failure of gravity over long
distances \cite{MOND}, but at present the evidence for new
gravitational physics at large distances does not seem compelling
\cite{DMvsNP}.) The perceived crisis is the absence of an
over-arching theoretical framework within which both successes can
be accommodated. Our brains are effectively split into two
incommunicative hemispheres, with quantum physics living in one
and classical general relativity in the other.

The absence of such a framework would indeed be a crisis for
theoretical physics, since real theoretical predictions are
necessarily approximate. Controllable results always require some
understanding of the size of the contributions being neglected in
any given calculation. If quantum effects in General Relativity
cannot be quantified, this must undermine our satisfaction with
the experimental success of its classical predictions.

It is the purpose of this article to present the modern point of
view on these issues, which has emerged since the early 1980's.
According to this point of view there is no such crisis, because
the problems of quantizing gravity within the experimentally
accessible situations are similar to those which arise in a host
of other non-gravitational applications throughout physics. As
such, the size of quantum corrections can be safely estimated and
are extremely small. The theoretical framework which allows this
quantification is the formalism of \emph{effective field
theories}, whose explanation makes up the better part of this
article. In so doing we shall see that although there can be
little doubt of the final outcome, the explicit determination of
the size of sub-leading quantum effects in gravity has in many
cases come only relatively recently, and a complete quantitative
analysis of the size of quantum corrections remains a work in
progress.

\subsection{Identifying Where the Problems Lie}
This is not to say that there are no challenging problems
remaining in reconciling quantum mechanics with gravity. On the
contrary, many of the most interesting issues remain to be solved,
including the identification of what the right observables should
be, and understanding how space and time might emerge from more
microscopic considerations. For the rest of the discussion it is
useful to separate these deep, unsolved issues of principle from
the more prosaic, technical problem of General Relativity's
\emph{non-renormalizability}.

There have been a number of heroic attempts to quantize gravity
along the lines of other field theories \cite{EarlyQG},
\cite{CanonQG}, \cite{LivRev-QG} and it was recognized early on
that General Relativity is not renormalizable. It is this
technical problem of non-renormalizability which in practice has
been the obstruction to performing quantum calculations with
General Relativity. As usually stated, the difficulty with
non-renormalizable theories is that they are not predictive, since
the obtention of well-defined predictions potentially requires an
infinite number of divergent renormalizations.

It is not the main point of the present review to recap the
techniques used when quantizing the gravitational field, nor to
describe in detail its renormalizability. Rather, this review is
intended to describe the modern picture of what renormalization
means, and why non-renormalizable theories need not preclude
making meaningful predictions. This point of view is now
well-established in many areas --- such as particle, nuclear and
condensed-matter physics --- where non-renormalizable theories
arise. In these other areas of physics predictions can be made
with non-renormalizable theories (including quantum corrections)
and the resulting predictions are well-verified experimentally.
The key to making these predictions is to recognize that they must
be made within the context of a low-energy expansion, in powers of
$E/M$ (energy divided by some heavy scale intrinsic to the
problem). Within the validity of this expansion theoretical
predictions are under complete control.

The lesson for quantum gravity is clear: non-renormalizability is
not in itself an obstruction to performing predictive quantum
calculations provided the low-energy nature of these predictions
in powers of $E/M$, for some $M$, is borne in mind. What plays the
role of the heavy scale $M$ in the case of quantum gravity? It is
tempting to identify this scale with the Planck mass, $M_p^{-2} =
8 \pi G$ (with $G$ denoting Newton's constant), and in some
circumstances this is the right choice. But as we shall see $M$
need not be $M_p$, and for some applications might instead be the
electron mass, $m_e$, or some other scale. One of the points of
quantifying the size of quantum corrections is to identify more
precisely what the important scales are for a given
quantum-gravity application.

Once it is understood how to use non-renormalizable theories, the
size of quantum effects can be quantified and it becomes clear
where the real problems of quantum gravity are pressing and where
they are not. In particular, the low-energy expansion proves to be
an extremely good approximation for all of the present
experimental tests of gravity, making quantum corrections
negligible for these tests. By contrast, the low-energy nature of
quantum-gravity predictions implies that quantum effects are
important where gravitational fields become very strong, such as
inside black holes or near cosmological singularities. This is
what makes the study of these situations so interesting: it is
through their study that progress on the more fundamental issues
of quantum gravity is likely to come.

\subsection{A Road Map}
The remainder of this article is organized in the following way.

\begin{itemize}
\item {\it Section 2: Effective Field Theories} does not involve
gravity at all, but instead first describes why effective field
theories are useful in other branches of physics. The discussion
is kept concrete by considering a simple toy model, for which it
is argued how some applications make it useful to keep track of
how small ratios of energy scales appear in physical observables.
In particular, considerable simplification can be achieved if an
expansion in small energy ratios is performed as early as possible
in the calculation of low-energy observables. The theoretical tool
for achieving this simplification is the effective lagrangian, and
its definition and use is briefly summarized using the toy model
as an explicit example.

\item {\it Section 3: Quantum Gravity as an Effective Theory}
describes how the tools of the previous section may be applied to
calculating quantum effects including the gravitational field. In
particular, it is shown how to make predictions despite General
Relativity's non-renormalizability, since effective lagrangians
are generically not renormalizable. As we shall see, however, some
of the main results one would like to have regarding the size of
quantum corrections to arbitrary loop orders remain incomplete.

\item {\it Section 4: Explicit Applications} of these ideas are
described in this section, which use the above results to compute
quantum corrections to several gravitational results for two kinds
of sources. These calculations compute the leading quantum
corrections to Newton's Law between two slowly-moving point
particles, and to the gravitational force between two cosmic
strings (both in 3+1 spacetime dimensions).

\item {\it Section 5: Conclusions} are briefly summarized in the final
section.

\end{itemize}



\section{Effective Field Theories}
\label{section:EFT}
This section describes the effective-lagrangian technique within
the context of a simple toy model, closely following the
discussion of ref.~\cite{Ode}.

In all branches of theoretical physics a key part of any good
prediction is a careful assessment of the theoretical error which
the prediction carries. Such an assessment is a precondition for
any detailed quantitative comparison with experiment. As is clear
from numerous examples throughout physics, this assessment of
error usually is usually reliably determined based on an
understanding of the small quantities which control the
corrections to the approximations used when making predictions.
Perhaps the most famous example of such a small quantity might be
the fine-structure constant, $\alpha = e^2/\hbar c$, which
quantifies the corrections to electromagnetic predictions of
elementary particle properties or atomic energy levels.

\subsection{The Utility of Low-Energy Approximations}
It sometimes happens that predictions are much more accurate than
would be expected based on an assessment of the approximations on
which they appear to be based. A famous example of this is
encountered in the precision tests of Quantum Electrodynamics,
where the value of the fine-structure constant, $\alpha$, was
until recently obtained using the Josephson effect in
superconductivity.

A DC potential difference applied at the boundary between two
superconductors can produce an AC Josephson current whose
frequency is precisely related to the size of the applied
potential and the electron's charge. Precision measurements of
frequency and voltage are in this way converted into a precise
measurement of $e/\hbar$, and so of $\alpha$. But use of this
effect to determine $\alpha$ only makes sense if the predicted
relationship between frequency and voltage is also known to an
accuracy which is better than the uncertainty in $\alpha$.

It is, at first sight, puzzling how such an accurate prediction
for this effect can be possible. After all, the prediction is made
within the BCS theory of superconductivity, \cite{BCS} which
ignores most of the mutual interactions of electrons, focussing
instead on a particular pairing interaction due to phonon
exchange. Radical though this approximation might appear to be,
the theory works rather well (in fact, surprisingly well), with
its predictions often agreeing with experiment to within several
percent. But expecting successful predictions with an accuracy of
parts per million or better would appear to be optimistic indeed!

The astounding theoretical accuracy required to successfully
predict the Josephson frequency may be understood at another
level, however. The key observation is that this prediction does
not rely at all on the details of the BCS theory, depending
instead only on the symmetry-breaking pattern which it predicts.
Once it is known that a superconductor spontaneously breaks the
$U(1)$ gauge symmetry of electromagnetism, the Josephson
prediction follows on general grounds in the low-energy
limit.\cite{WeinbergSC} The validity of the prediction is
therefore not controlled by the approximations made in the BCS
theory, since {\it any} theory with the same low-energy
symmetry-breaking pattern shares the same predictions.

The accuracy of the predictions for the Josephson effect are
therefore founded on symmetry arguments, and on the validity of a
low-energy approximation. Quantitatively, the low-energy
approximation involves the neglect of powers of the ratio of two
scales, $\omega/\Omega$, where $\omega$ is the low energy scale of
the observable under consideration --- like the applied voltage in
the Josephson effect --- and $\Omega$ is the higher energy scale
--- such as the superconducting gap energy --- which is intrinsic
to the system under study.

Indeed, arguments based on a similar low-energy approximation may
also be used to explain the surprising accuracy of many other
successful models throughout physics, including the BCS theory
itself.\cite{Polchinski,Shankar,Frohlich} This is accomplished by
showing that only the specific interactions used by the BCS theory
are relevant at low energies, with all others being suppressed in
their effects by powers of a small energy ratio.

Although many of these arguments were undoubtedly known in various
forms by the experts in various fields since very early days, the
systematic development of these arguments into precision
calculational techniques has happened more recently. With this
development has come considerable cross-fertilization of
techniques between disciplines, with the realization that the same
methods play a role across diverse disciplines within physics.

The remainder of this lecture briefly summarizes the techniques
which have been developed to exploit low-energy approximations.
These are most efficiently expressed using effective-lagrangian
methods, which are designed to take advantage of the simplicity of
the low-energy limit as early as possible within a calculation.
The gain in simplicity so obtained can be the decisive difference
between a calculation's being feasible rather than being too
difficult to entertain.

Besides providing this kind of practical advantage,
effective-lagrangian techniques also bring real conceptual
benefits because of the clear separation they permit between of
the effects of different scales. Both of these kinds of advantages
are illustrated here using explicit examples. First \S2\ presents
a toy model involving two spinless particles to illustrate the
general method, as well as some of its calculational advantages.
This is followed by a short discussion of the conceptual
advantages, with quantum corrections to classical general
relativity, and the associated problem of the nonrenormalizability
of gravity, taken as the illustrative example.

\subsection{A Toy Example}
In order to make the discussion as concrete as possible, consider
the following model for a single complex scalar field, $\phi$:
\begin{equation}
\label{abeltoymodel}
 {\cal L} = - \partial_\mu \phi^*
 \partial^\mu \phi - V(\phi^* \phi) \,,
\end{equation}
with
\begin{equation}
 V = {\lambda^2 \over 4} \; \left( \phi^* \phi - v^2 \right)^2 \,.
\end{equation}
This theory enjoys a continuous $U(1)$ symmetry of the form $\phi
\to e^{i\omega} \; \phi$, where the parameter, $\omega$, is a
constant. The two parameters of the model are $\lambda$ and $v$.
Since $v$ is the only dimensionful quantity it sets the model's
overall energy scale.

The semiclassical approximation is justified if the dimensionless
quantity $\lambda$ should be sufficiently small. In this
approximation the vacuum field configuration is found by
minimizing the system's energy density, and so is given (up to a
$U(1)$ transformation) by $\phi = v$. For small $\lambda$ the
spectrum consists of two weakly-interacting particle types
described by the fields ${\cal R}$ and ${\cal I}$, where $\phi =
\left( v + \frac{1}{\sqrt2} \; {\cal R} \right) + \frac{i}{\sqrt2}
\; {\cal I}$. To leading order in $\lambda$ the particle masses
are $m_{\scriptscriptstyle I} = 0$ and $m_{\scriptscriptstyle R} =
\lambda v$.

The low-energy regime in this model is $E \ll
m_{\scriptscriptstyle R}$. The masslessness of ${\cal I}$ ensures
the existence of degrees of freedom in this regime, with the
potential for nontrivial low-energy interactions, which we next
explore.

\subsubsection{Massless-Particle Scattering}
The interactions amongst the particles in this model are given by
the scalar potential :
\begin{equation}
\label{potl} V = {\lambda^2 \over 16} \; \Bigl( 2 \sqrt2 \, v
{\cal R} + {\cal R}^2 + {\cal I}^2 \Bigr)^2.
\end{equation}

\begin{figure}[t]
  \def\epsfsize#1#2{0.8#1}
  \centerline{\epsfbox{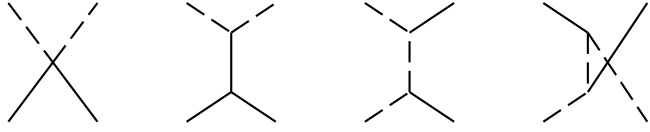}}
  \caption{\it The Feynman graphs responsible for tree-level
  ${\cal R}-{\cal I}$
  scattering in the toy model. Here solid lines denote ${\cal R}$
  particles and dashed lines represent ${\cal I}$ particles.}
  \label{figure:scattering}
\end{figure}

Imagine using the potential of eq.~(\ref{potl}) to calculate the
amplitude for the scattering of ${\cal I}$ particles at low
energies to lowest-order in $\lambda$. For example, the Feynman
graphs describing tree-level ${\cal I}-{\cal R}$ scattering are
given in Fig.~(\ref{figure:scattering}). The $S$-matrix obtained
by evaluating the analogous tree-level diagrams for ${\cal I}$
self-scattering is proportional to the following invariant
amplitude:
\begin{eqnarray}
\label{smatrixresult}
 {\cal A} &=& - \; {3 \lambda^2 \over 2} +
 \left( { \lambda^2 v \over \sqrt2}
 \right)^2 \left[ {1 \over (s+r)^2 + m_{\scriptscriptstyle R}^2
 - i\epsilon}  \right. \nonumber\\
 && \qquad\qquad \left. + {1 \over (r-r')^2 + m_{\scriptscriptstyle
 R}^2 -i \epsilon} + {1 \over (r - s')^2 + m_{\scriptscriptstyle
 R}^2 - i\epsilon}\right],
\end{eqnarray}
where $s^\mu$ and $r^\mu$ (and $s'{}^\mu$ and $r'{}^\mu$) are the
4-momenta of the initial (and final) particles.

An interesting feature of this amplitude is that when it is
expanded in powers of external four-momenta, both its leading and
next-to-leading terms vanish. That is:
\begin{eqnarray}
\label{zeromomlim}
 {\cal A} &=&  \left[ - \; {3 \lambda^2 \over 2}
 + {3 \over m_{\scriptscriptstyle R}^2} \, \left( { \lambda^2 v
 \over \sqrt2} \right)^2 \right] + {2 \over m_{\scriptscriptstyle
 R}^4} \; \left( { \lambda^2 v \over \sqrt2} \right)^2 \Bigl[- r
 \cdot s
 + r \cdot r' + r \cdot s' \Bigr]  \nonumber\\
 && \qquad\qquad\qquad\qquad \qquad\qquad\qquad
 + \, O(\hbox{quartic in momenta}) \nonumber\\
 &=& 0 + O(\hbox{quartic in momenta}) .
\end{eqnarray}
The last equality uses conservation of 4-momentum: $s^\mu + r^\mu
= s'{}^\mu + r'{}^\mu$ and the massless mass-shell condition $r^2
= 0$. Something similar occurs for ${\cal R}-{\cal I}$ scattering,
which also vanishes due to a cancellation amongst the graphs of
Fig.~(\ref{figure:scattering}) in the zero-momentum limit.

Clearly the low-energy particles interact more weakly than would
be expected given a cursory inspection of the scalar potential,
eq.~(\ref{potl}), since at tree level the low-energy scattering
rate is suppressed by at least eight powers of the small energy
ratio $r = E/m_{\scriptscriptstyle R}$. The real size of the
scattering rate might depend crucially on the relative size of $r$
and $\lambda^2$, should the vanishing of the leading low-energy
terms turn out to be an artifact of leading-order perturbation
theory.

If ${\cal I}$ scattering were of direct experimental interest, one
can imagine considerable effort being invested in obtaining
higher-order corrections to this low-energy result. And the final
result proves to be quite interesting: as may be verified by
explicit calculation, the first two terms in the low-energy
expansion of ${\cal A}$ vanish order-by-order in perturbation
theory. Furthermore, a similar suppression turns out also to hold
for all other amplitudes involving ${\cal I}$ particles, with the
$n$-point amplitude for ${\cal I}$ scattering being suppressed by
$n$ powers of $r$.

Clearly the hard way to understand these low-energy results is to
first compute to all orders in $\lambda$ and then expand the
result in powers of $r$. A much more efficient approach exploits
the simplicity of small $r$ {\sl before} calculating scattering
amplitudes.

\subsection{The Toy Model Revisited}
The key to understanding this model's low-energy limit is to
recognize that the low-energy suppression of scattering amplitudes
(as well as the exact massless of the light particle) is a
consequence of the theory's $U(1)$ symmetry. (The massless state
has these properties because it is this symmetry's Nambu-Goldstone
boson.\cite{ChiPT,physica,GBreviews,Burgess00}) The simplicity of
the low-energy behaviour is therefore best displayed by:

\begin{enumerate}
\item
Making the symmetry explicit for the low-energy degrees of
freedom;
\item
Performing the low-energy approximation as early as possible.
\end{enumerate}

\subsubsection{Exhibiting the Symmetry}

The $U(1)$ symmetry can be made to act exclusively on the field
which represents the light particle by parameterizing the theory
using a different set of variables than ${\cal I}$ and ${\cal R}$.
To this end imagine instead using polar coordinates in field
space:
\begin{equation}
\label{polcoords} \phi(x) = \chi(x) \; e^{i \theta(x)} .
\end{equation}
In terms of $\theta$ and $\chi$ the action of the $U(1)$ symmetry
is simply $\theta \to \theta + \omega$, and the model's Lagrangian
becomes:
\begin{equation}
\label{linpolcoords} {\cal L} = - \partial_\mu \chi \partial^\mu
\chi - \chi^2 \partial_\mu \theta
\partial^\mu \theta - V(\chi^2).
\end{equation}
The semiclassical spectrum of this theory is found by expanding
${\cal L}$ in powers of the canonically-normalized fluctuations,
$\chi' = \sqrt2 (\chi - v)$ and $\theta' = \sqrt2 \, v \, \theta$,
about the vacuum $\chi = v$, revealing that $\chi'$ describes the
mass-$m_{\scriptscriptstyle R}$ particle while $\theta'$
represents the massless particle.

With the $U(1)$ symmetry realized purely on the massless field,
$\theta$, we may expect good things to happen if we identify the
low-energy dynamics.

\subsubsection{Timely Performance the Low-Energy Approximation}
To properly exploit the symmetry of the low-energy limit we
integrate out all of the high-energy degrees of freedom as the
very first step, leaving the inclusion of the low-energy degrees
of freedom to last. This is done most efficiently by computing the
following low-energy effective (or, Wilson) action.

A conceptually simple (but cumbersome in practice) way to split
degrees of freedom into `heavy' and `light' categories is to
classify all field modes in momentum space as heavy if (in
Euclidean signature) they satisfy $p^2 + m^2 > \Lambda^2$ where
$m$ is the corresponding particle mass and $\Lambda$ is an
appropriately chosen cutoff.

Light modes are then all of those which are not heavy. The cutoff,
$\Lambda$, which defines the boundary between these two kinds of
modes is chosen to lie well below the high-energy scale ({\it
i.e.}\ well below $m_{\scriptscriptstyle R}$ in the toy model) but
is also chosen to lie well above the low-energy scale of ultimate
interest (like the centre-of-mass energies, $E$, of low-energy
scattering amplitudes). Notice that in the toy model the heavy
degrees of freedom defined by this split include all modes of the
field $\chi'$, as well as the high-frequency components of the
massless field $\theta'$.

If $h$ and $\ell$ schematically denote the fields which are,
respectively, heavy or light in this characterization, then the
influence of heavy fields on light-particle scattering at low
energies is completely encoded in the following effective
lagrangian:
\begin{equation}
\label{leffdef} \exp\left[i \int d^4x \; {\cal L}_{\rm
eff}(\ell,\Lambda) \right] = \int {\cal D}h_\Lambda \; \exp\left[
i \int d^4x \; {\cal L}(\ell,h) \right].
\end{equation}
The $\Lambda$-dependence which is introduced by the
low-energy/high-energy split of the integration measure is
indicated explicitly in this equation.

Physical observables at low energies are now computed by
performing the remaining path integral over the light degrees of
freedom only. By virtue of its definition, each configuration in
the integration over light fields is weighted by a factor of
$\exp\left[i \int d^4x \; {\cal L}_{\rm eff}(\ell) \right]$
implying that the effective lagrangian weights the low-energy
amplitudes in precisely the same way as the classical lagrangian
does for the integral over both heavy and light degrees of
freedom. In detail, the effects of virtual contributions of heavy
states appear within the low-energy theory through the
contributions of new effective interactions, such as

Although this kind of low-energy/high-energy split in terms of
cutoffs most simply illustrates the conceptual points of interest,
in practical calculations it is usually dimensional regularization
which is more useful. This is particularly true for theories (like
General Relativity) involving gauge symmetries, which can be
conveniently kept manifest using dimensional regularization. We
therefore return to this point in subsequent sections to explain
how dimensional regularization can be used with an effective field
theory.

\subsubsection{Implications for the Low-Energy Limit}
Now comes the main point. When applied to the toy model the
condition of symmetry and the restriction to the low-energy limit
together have strong implications for ${\cal L}_{\rm
eff}(\theta)$. Specifically:

\begin{enumerate}
\item
Invariance of ${\cal L}_{\rm eff}(\theta)$ under the symmetry
$\theta \to \theta + \omega$ implies ${\cal L}_{\rm eff}$ can
depend on $\theta$ only through the invariant quantity
$\partial_\mu \theta$.
\item
Interest in the low-energy limit permits the expansion of ${\cal
L}_{\rm eff}$ in powers of derivatives of $\theta$. Because only
low-energy functional integrals remain to be performed, higher
powers of $\partial_\mu\theta$ correspond in a calculable way to
higher suppression of observables by powers of
$E/m_{\scriptscriptstyle R}$.
\end{enumerate}

Combining these two observations leads to the following form for
${\cal L}_{\rm eff}$:
\begin{eqnarray}
\label{Leffform}
 {\cal L}_{\rm eff} &=& - v^2 \; \partial_\mu
 \theta \, \partial^\mu \theta + a \, (\partial_\mu \theta \,
 \partial^\mu \theta)^2 + {b \over m_{\scriptscriptstyle R}^2} \; (\partial_\mu
 \theta \, \partial^\mu \theta)^3
 \nonumber\\
 && \qquad\qquad\qquad\qquad + {c \over m_{\scriptscriptstyle R}^2}
 \; (\partial_\mu \theta \, \partial^\mu \theta)
 \partial_\lambda \partial^\lambda
 (\partial_\nu \theta \, \partial^\nu \theta) + \cdots,
\end{eqnarray}
where the ellipses represent terms which involve more than six
derivatives, and so more than two inverse powers of
$m_{\scriptscriptstyle R}$.

A straightforward calculation confirms the form,
eq.~(\ref{Leffform}), in perturbation theory, but with the
additional information
\begin{equation}
\label{treecoeffs} a_{\rm pert} = {1 \over 4 \lambda^2} +
O(\lambda^0), \qquad b_{\rm pert} = - \; {1 \over 4 \lambda^2} +
O(\lambda^0), \qquad c_{\rm pert} = {1 \over 4 \lambda^2}  +
O(\lambda^0).
\end{equation}

In this formulation it is clear that each additional factor of
$\theta$ is always accompanied by a derivative, and so implies an
additional power of $r$ in its contribution to all light-particle
scattering amplitudes. Because eq.~(\ref{Leffform}) is derived
assuming only general properties of the low energy effective
lagrangian, its consequences (such as the suppression by $r^n$ of
low-energy $n$-point amplitudes) are insensitive of the details of
underlying model. They apply, in particular, to all orders in
$\lambda$.

Conversely, the details of the underlying physics only enter
through specific predictions, such as eqs.~(\ref{treecoeffs}), for
the low-energy coefficients $a,b$ and $c$. Different models having
a $U(1)$ Goldstone boson in their low-energy spectrum can differ
in the low-energy self-interactions of this particle only through
the values they predict for these coefficients.

\subsubsection{Redundant Interactions}

The effective lagrangian, (\ref{Leffform}), does not contain all
possible polynomials of $\partial_\mu \theta$. For example, two
terms involving 4 derivatives which are not written are
\begin{equation} \label{redundant}
 {\cal L}_{\rm redundant} = d \; \Box \theta \, \Box \theta + e\;
 \partial_\mu \theta \, \Box \partial^\mu \theta \,,
\end{equation}
where $d$ and $e$ are arbitrary real constants. These terms are
omitted because their inclusion would not alter any of the
predictions of ${\cal L}_{\rm eff}$. Because of this, interactions
such as those in eq.~(\ref{redundant}) are known as
\emph{redundant} interactions.

There are two reasons why such terms do not contribute to physical
observables. The first reason is the old saw that states that
total derivatives may be dropped from an action. More precisely,
such terms may be integrated to give either topological
contributions or surface terms evaluated at the system's boundary.
They may therefore be dropped provided that none of the physics of
interest depends on the topology or what happens on the system's
boundaries. (See, however, ref.~\cite{Casimir} and references
therein for a concrete example where boundary effects play an
important role within an effective field theory.) Certainly
boundary terms are irrelevant to the form of the classical field
equations far from the boundary. They also do not contribute
perturbatively to scattering amplitudes, as may be seen from the
Feynman rules which are obtained from a simple total derivative
interaction like
\begin{equation}
 \Delta {\cal L} = g\,  \partial_\mu (\partial^\mu \theta \Box \theta) =
 g \Bigl( \Box \theta \Box \theta + \partial^\mu \theta \Box \partial_\mu
 \theta \Bigr) \, ,
\end{equation}
since these are proportional to
\begin{equation}
 g \, (p^2 q^2 + p^\mu q_\mu q^2) \, \delta^4(p+q) =
 g \, q^2 p^\mu (p_\mu + q_\mu) \, \delta^4(p+q) =  0 \,.
\end{equation}
This shows that the two interactions of eq.~(\ref{redundant}) are
not independent, since we can integrate by parts to replace the
couplings $(d,e)$ with $(d',e') = (d-e,0)$.

The second reason why interactions might be physically irrelevant
(and so redundant) is if they may be removed by performing a field
redefinition. For instance under the infinitesimal redefinition
$\delta \theta  = A \, \Box \theta $, the leading term in the
low-energy action transforms to
\begin{equation}
 \delta (- v^2 \, \partial_\mu \theta \partial^\mu \theta) = - 2A
 v^2 \, \partial_\mu \theta \Box \partial^\mu \theta \,.
\end{equation}
This redefinition can be used to set the effective coupling $e$ to
zero, simply by choosing $2Av^2 = e$. This argument can be
repeated order-by-order in powers of $1/m_{\scriptscriptstyle R}$
to remove more and more terms in ${\cal L}_{\rm eff}$ without
affecting physical observables.

Since the variation of the lowest-order action is always
proportional to its equations of motion, it is possible to remove
in this way \emph{any} interaction which vanishes when evaluated
at the solution to the lower-order equations of motion. Of course,
a certain amount of care must be used when so doing. For instance,
if our interest is in how the $\theta$ field affects the
interaction energy of classical sources, we must add a source
coupling, $\Delta {\cal L} = J^\mu
\partial_\mu\theta$ to the lagrangian. Once this is done the
lowest-order equations of motion become $2 v^2 \Box \theta =
\partial_\mu J^\mu$, and so an effective interaction like $\Box
\theta \Box \theta$ is no longer completely redundant. It is
instead equivalent to the contact interactions like $(\partial_\mu
J^\mu)^2/(4v^4)$.

\subsection{Lessons Learned}
It is clear that the kind of discussion given for the toy model
can be performed equally well for any other system having two
well-separated energy scales. There are a number of features of
this example which also generalize to these other systems. It is
the purpose of this section to briefly list some of these
features.

\subsubsection{Why are Effective Lagrangians not More Complicated?}
${\cal L}_{\rm eff}$ as computed in the toy model is not a
completely arbitrary functional of its argument, $\theta$. For
example, ${\cal L}_{\rm eff}$ is {\sl real} and not complex, and
it is {\sl local} in the sense that (to any finite order in
$1/m_{\scriptscriptstyle R}$) it consists of a finite sum of
powers of the field $\theta$ and its derivatives, all evaluated at
the same point.

Why should this be so? Both of these turn out to be general
features (so long as only massive degrees of freedom are
integrated out) which are inherited from properties of the
underlying physics at higher energies.

\begin{itemize}
\item[{\it (i)}]
{\it Reality:} The reality of ${\cal L}_{\rm eff}$ is a
consequence of the unitarity of the underlying theory, and the
observation that the degrees of freedom which are integrated out
to obtain ${\cal L}_{\rm eff}$ are excluded purely on the grounds
of their energy. As a result, if no heavy degrees of freedom
appear as part of an initial state, energy conservation precludes
their being produced by scattering and so appearing in the final
state.

Since ${\cal L}_{\rm eff}$ is constructed to reproduce this time
evolution of the full theory, it must be real in order to give a
hermitian Hamiltonian as is required by unitary time
evolution.\footnote{There can be circumstances for which energy is
not the criterion used to define the effective theory, and for
which ${\cal L}_{\rm eff}$ is not real. The resulting failure of
unitarity in the effective theory reflects the possibility in
these theories of having states in the effective theory converting
into states that have been removed in its definition.}

\item[{\it (ii)}]
{\it Locality:} The locality of ${\cal L}_{\rm eff}$ is also a
consequence of excluding high-energy states in its definition,
together with the Heisenberg Uncertainty Relations. Although
energy and momentum conservation preclude the direct production of
heavy particles (like those described by $\chi$ in the toy model)
from an initial low-energy particle configuration, it does not
preclude their {\it virtual} production.

That is, heavy particles may be produced so long as they are then
re-destroyed sufficiently quickly. Such virtual production is
possible because the Uncertainty Relations permit energy not to be
precisely conserved for states which do not live indefinitely
long. A virtual state whose production requires energy
nonconservation of order $\Delta E \sim M$ therefore cannot live
longer than $\Delta t \sim 1/M$, and so its influence must appear
as being local in time when observed only with probes having much
smaller energy. Similar arguments imply locality in space for
momentum-conserving systems.

Since it is the mass $M$ of the heavy particle which sets the
scale over which locality applies once it is integrated out, it is
$1/M$ which appears with derivatives of low-energy fields when
${\cal L}_{\rm eff}$ is written in a derivative expansion.

\end{itemize}

\subsection{Predictiveness and Power Counting}
The entire rationale of an effective lagrangian is to incorporate
the virtual effects of high-energy particles in low-energy
processes, order-by-order in powers of the small ratio, $r$, of
these two scales ({\it e.g.} $r = E/m_{\scriptscriptstyle R}$ in
the toy model). In order to use an effective lagrangian it is
therefore necessary to know which terms contribute to physical
processes to any given order in $r$.

This determination is explicitly possible if the low-energy
degrees of freedom are weakly interacting, because in this case
perturbation theory in the weak interactions may be analyzed
graphically, permitting the use of power-counting arguments to
systematically determine where powers of $r$ originate. Notice
that the assumption of a weakly-interacting low-energy theory does
{\it not} presuppose the underlying physics to be also weakly
interacting. For instance, for the toy model the Goldstone boson
of the low-energy theory is weakly interacting provided only that
the $U(1)$ symmetry is spontaneously broken, since its
interactions are all suppressed by powers of $r$. Notice that this
is true {\it independent} of the size of coupling, $\lambda$, of
the underlying theory.

For example, in the toy model the effective lagrangian takes the
general form:
\begin{equation} \label{genformLeff}
 {\cal L}_{\rm eff} = v^2 m_{\scriptscriptstyle
 R}^2 \sum_{id} {c_{id} \over m_{\scriptscriptstyle R}^d} \; {\cal
 O}_{id},
\end{equation}
where the sum is over interactions, ${\cal O}_{id}$, involving $i$
powers of the dimensionless field $\theta$ and $d$ derivatives.
The power of $m_{\scriptscriptstyle R}$ premultiplying each term
is chosen to ensure that the coefficient $c_{id}$ is
dimensionless. (For instance, the interaction $(\partial_\mu
\theta \,
\partial^\mu \theta)^2$ has $i = d = 4$.) There are two useful
properties which all of the operators in this sum must satisfy:
\begin{enumerate}
\item
$d$ must be even by virtue of Lorentz invariance.
\item
Since the sum is only over interactions, it does not include the
kinetic term, which is the unique term for which $d=i=2$.
\item
The $U(1)$ symmetry implies every factor of $\theta$ is
differentiated at least once, and so $d \ge i$. Furthermore, any
term linear in $\theta$ must therefore be a total derivative, and
so may be omitted, implying $i \ge 2$ without loss.
\end{enumerate}

\subsubsection{Power-Counting Low-Energy Feynman Graphs}
It is straightforward to track the powers of $v$ and
$m_{\scriptscriptstyle R}$ that interactions of the form
(\ref{genformLeff}) contribute to an $\ell$-loop contribution to
the amplitude, ${\cal A}_{\cal E}(E)$, for the scattering of $N_i$
initial ${\cal E}$ Goldstone bosons into $N_f$ final Goldstone
bosons at centre-of-mass energy $E$. The label ${\cal E} = N_i +
N_f$ here denotes the number of external lines in the
corresponding graph. (The steps presented in this section closely
follow the discussion of ref.~ \cite{Burgess00}.)

With the desire of also being able to include later examples,
consider the following slight generalization of the lagrangian of
(\ref{genformLeff}):
\begin{equation} \label{leffpc}
 {\cal L}_{\rm eff} = f^4 \sum_{n} {c_{n} \over M^{d_n}} \;
 {\cal O}_n \left( {\phi \over v} \right) \,.
\end{equation}
Here $\phi$ denotes a generic boson field, $c_n$ are again
dimensionless coupling constants which we imagine to be at most
$O(1)$, and $f, M$ and $v$ are mass scales of the underlying
problem. The sum is again over operators which are powers of the
fields and their derivatives, and $d_n$ is the dimension of the
operator ${\cal O}_n$, in powers of mass. For example, in the
toy-model application we have $f^2 = v \, m_{\scriptscriptstyle
R}$, $M = m_{\scriptscriptstyle R}$ and we have written $\theta =
\phi/v$. In the toy-model example the sum over $n$ corresponds to
the sum over $i$ and $d$ and $d_n = d$.

Imagine using this lagrangian to compute a scattering amplitude,
${\cal A}_{\cal E}(E)$, involving the scattering of ${\cal E}$
relativistic particles whose energy and momenta are of order $E$.
We wish to focus on the contribution to ${\cal A}$ due to a
Feynman graph having $I$ internal lines and $V_{ik}$ vertices. The
labels $i$ and $k$ here indicate two characteristics of the
vertices: $i$ counts the number of lines which converge at the
vertex, and $k$ counts the power of momentum which appears in the
vertex. Equivalently, $i$ counts the number of powers of the
fields, $\phi$, which appear in the corresponding interaction term
in the lagrangian, and $k$ counts the number of derivatives of
these fields which appear there.


\medskip\noindent {\it Some Useful Identities}

\medskip\noindent The positive integers, $I$, $E$ and $V_{ik}$, which
characterize the Feynman graph in question are not all independent
since they are related by the rules for constructing graphs from
lines and vertices.

The first such relation can be obtained by equating two equivalent
ways of counting how internal and external lines can end in a
graph. On the one hand, since all lines end at a vertex, the
number of ends is given by summing over all of the ends which
appear in all of the vertices: $\sum_{ik} i \, V_{ik}$. On the
other hand, there are two ends for each internal line, and one end
for each external line in the graph: $2 I + E$. Equating these
gives the identity which expresses the `conservation of ends':
\begin{equation} \label{consofends}
 2 I + E = \sum_{ik} i \,  V_{ik}, \qquad
 \hbox{(Conservation of Ends)}.
\end{equation}

A second useful identity gives the number of loops, $L$, for each
(connected) graph:
\begin{equation} \label{loopdef}
 L = 1 + I - \sum_{ik} V_{ik}, \qquad
 \hbox{(Definition of $L$)}.
\end{equation}
For simple planar graphs, this last equation agrees with the
intuitive notion of what the number of loops in a graph means,
since it expresses a topological invariant which states how the
Euler number for a disc can be expressed in terms of the number of
edges, corners and faces of the triangles in one of its
triangularization. For graphs which do not have the topology of a
plane, eq.~(\ref{loopdef}) should instead be read as
\emph{defining} the number of loops.


\medskip\noindent {\it Dimensional Estimates}

\medskip\noindent
We now collect the dependence of ${\cal A}_{\cal E}(a)$ on the
parameters appearing in ${\cal L}_{\rm eff}$. Reading the Feynman
rules from the lagrangian of eq.~(\ref{leffpc}) shows that the
vertices in the Feynman graph contribute the following factor:
\begin{equation} \label{vertexcont}
 \hbox{(Vertex)} =  \prod_{ik} \left[ i (2
 \pi)^4 \delta^4(p) \; \left( {p \over M} \right)^k \; \left( {f^4
 \over v^i} \right) \right]^{V_{ik}},
\end{equation}
where $p$ generically denotes the various momenta running through
the vertex. Similarly, there are $I$ internal lines in the graph,
each of which contributes the additional factor:
\begin{equation} \label{internallinecont}
 \hbox{(Internal Line)} = \left[ -i
 \int {d^4 p \over (2 \pi)^4} \; \left( {M^2 v^2 \over f^4} \right)
 \; {1 \over p^2 + m^2} \right],
\end{equation}
where, again, $p$ denotes the generic momentum flowing through the
line. $m$ generically denotes the mass of any light particles
which appear in the effective theory, and it is assumed that the
kinetic terms which define their propagation are those terms in
${\cal L}_{\rm eff}$ involving two derivatives and two powers of
the fields, $\phi$.

As usual for a connected graph, all but one of the
momentum-conserving delta functions in eq.~(\ref{vertexcont}) can
be used to perform one of the momentum integrals in
eq.~(\ref{internallinecont}). The one remaining delta function
which is left after doing so depends only on the external momenta,
$\delta^4(q)$, and expresses the overall conservation of
four-momentum for the process. Future formulae are less cluttered
if this factor is extracted once and for all, by defining the
reduced amplitude, $\tilde {\cal A}$, by
\begin{equation} \label{redampdef}
 {\cal A}_{\cal E}(E) = i (2 \pi)^4 \delta^4(q) \;
 \tilde {\cal A}_{\cal E}(E).
\end{equation}
Here $q$ generically represents the external four-momenta of the
process, whose components are of order $E$ in size.

The number of four-momentum integrations which are left after
having used all of the momentum-conserving delta functions is then
$I - \sum_{ik} V_{ik} + 1 = L$. This last equality uses the
definition, eq.~(\ref{loopdef}), of the number of loops, $L$.

We now estimate the result of performing the integration over the
internal momenta. To do so it is most convenient to regulate the
ultraviolet divergences which arise using dimensional
regularization.\footnote{We return below to a discussion of how
effective lagrangians can be defined using dimensional
regularization.} For dimensionally-regularized integrals, the key
observation is that the size of the result  is set on dimensional
grounds by the light masses or external momenta of the theory.
That is, if all external energies, $q$, are comparable to (or
larger than) the masses, $m$, of the light particles whose
scattering is being calculated, then $q$ is the light scale
controlling the size of the momentum integrations, so dimensional
analysis implies that an estimate of the size of the momentum
integrations is:
\begin{equation} \label{newdimgrounds}
 \int \cdots \int \left( {d^n p\over (2
 \pi)^n} \right)^A \; {p^B \over (p^2 + q^2)^C }  \sim \left( {1
 \over 4 \pi} \right)^{2A} q^{nA + B - 2C} ,
\end{equation}
with a dimensionless pre-factor which depends on the dimension,
$n$, of spacetime, and which may be singular in the limit that  $n
\to 4$. Notice that the assumption that $q$ is the largest
relevant scale in the low-energy theory explicitly excludes the
case of the scattering of non-relativistic particles.

One might worry whether such a simple dimensional argument can
really capture the asymptotic dependence of a complicated
multi-dimensional integral whose integrand is rife with potential
singularities. The ultimate justification for this estimate lies
with general results, like Weinberg's theorem
\cite{WeinbergsTheorem}, which underly the power-counting analyses
of renormalizability. These theorems ensure that the simple
dimensional estimates capture the correct behaviour up to
logarithms of the ratios of high- and low-energy mass scales.

With this estimate for the size of the momentum integrations, we
find the following contribution to the amplitude $\tilde {\cal
A}_{\cal E}(E)$:
\begin{equation} \label{intcontribution}
 \int \cdots \int \left( {d^4 p\over
 (2 \pi)^4} \right)^L \; {p^{X} \over (p^2 +
 q^2)^I } \sim \left( {1 \over 4 \pi} \right)^{2L} q^{Y} ,
\end{equation}
where $X = \sum_{ik} k V_{ik}$ and $Y = 4L - 2I + \sum_{ik} k
V_{ik}$. Liberal use of the identities (\ref{consofends}) and
(\ref{loopdef}), and taking $q \sim E$, allows this to be
rewritten as the following estimate:
\begin{equation} \label{aedwdimreg}
 \tilde {\cal A}_{\cal E}(E) \sim f^4 \; \left( {1
 \over v} \right)^{\cal E} \; \left( {M^2 \over 4 \pi f^2} \right)^{2L} \;
 \left( {E \over M} \right)^{P} \,,
\end{equation}
with $P = 2 + 2L + \sum_{ik} (k - 2) V_{ik}$. Equivalently, if we
group terms depending on $L$, eq.~(\ref{aedwdimreg}) may also be
written
\begin{equation} \label{aedwdimrega}
 \tilde {\cal A}_{\cal E}(E) \sim f^4 \; \left( {1
 \over v} \right)^{\cal E} \; \left( {ME \over 4 \pi f^2} \right)^{2L} \;
 \left( {E \over M} \right)^{P'} \,,
\end{equation}
with $P' = 2 + \sum_{ik} (k - 2) V_{ik}$.

Eq.~(\ref{aedwdimreg}) is the principal result of this section.
Its utility lies in the fact that it links the contributions of
the various effective interactions in the effective lagrangian,
(\ref{leffpc}), with the dependence of observables on small energy
ratios such as $r = E/M$. As a result it permits the determination
of which interactions in the effective lagrangian are required to
reproduce any given order in $E/M$ in physical observables.

Most importantly, eq.~(\ref{aedwdimreg}) shows how  to calculate
using nonrenormalizable theories. It implies that even though the
lagrangian can contain arbitrarily many terms, and so potentially
arbitrarily many  coupling constants, it is nonetheless predictive
{\it so long as  its predictions are only made for low-energy
processes, for which $E/M \ll 1$}.  (Notice also that the factor
$(M/f)^{4L}$ in eq.~(\ref{aedwdimreg}) implies,  all other things
being equal, the scale $f$ cannot be taken to be systematically
smaller than $M$ without ruining the validity of the loop
expansion in the effective low-energy theory. )

\subsubsection{Application to the Toy Model}
We now apply this power-counting estimate to the toy model
discussed earlier. Using the relations $f^2 = v \,
m_{\scriptscriptstyle R}$ and $M = m_{\scriptscriptstyle R}$ we
have
\begin{equation} \label{PCresult1}
 {\cal A}_{\cal E}(E) \sim v^2
 m_{\scriptscriptstyle R}^2 \left( {1 \over v} \right)^{\cal E}
 \left( { m_{\scriptscriptstyle R} \over 4 \pi v} \right)^{2L}
 \left( {E \over m_{\scriptscriptstyle R}} \right)^P,
\end{equation}
where $P = 2 + 2L + \sum_{id} (d-2) V_{id}$. As above, $V_{id}$
counts the number of times an interaction involving $i$ powers of
fields and $d$ derivatives appears in the amplitude. An equivalent
form for this expression is
\begin{equation} \label{PCresult2}
 {\cal A}_{\cal E}(E) \sim v^2
 E^2 \left( {1 \over v} \right)^{\cal E}
 \left( { E \over 4 \pi v} \right)^{2L} \prod_{i} \prod_{d>2}
 \left( {E \over m_{\scriptscriptstyle R}} \right)^{(d-2)V_{id}}
 \,.
\end{equation}

Eqs.~(\ref{PCresult1}) and (\ref{PCresult2}) have several
noteworthy features:
\begin{itemize}
\item
The condition $d \ge i \ge 2$ ensures that all of the powers of
$E$ appearing in ${\cal A}_{{\cal E}}$ are positive. Furthermore,
since $d = 2$ only occurs for the kinetic term, $d \ge 4$ for all
interactions, implying ${\cal A}_{\cal E}$ is suppressed by at
least 4 powers of $E$, and higher loops cost additional powers of
$E$. Furthermore, it is straightforward to identify the graphs
which contribute to ${\cal A}_{\cal E}$ to any fixed power of $E$.
\item
As is most clear from eq.~(\ref{PCresult2}), perturbation theory
in the effective theory relies only on the assumptions $E \ll 4
\pi v$ and $E \ll m_{\scriptscriptstyle R}$. In particular, it
does \emph{not} rely on the ratio $(m_{\scriptscriptstyle R}/4\pi
v)$ being small. Since $m_{\scriptscriptstyle R} = \lambda v$ in
the underlying toy model, this ratio is simply of order
$(\lambda/4 \pi )$, showing how weak coupling for the Goldstone
boson is completely independent of the strength of the couplings
in the underlying theory.
\end{itemize}

To see how eqs.~(\ref{PCresult1}) and (\ref{PCresult2}) are used,
consider the first few powers of $E$ in the toy model. For any
${\cal E}$ the leading contributions for small $E$ come from tree
graphs, $L = 0$. The tree graphs that dominate are those for which
$\sum_{id}' (d-2)V_{id}$ takes the smallest possible value. For
example, for 2-particle scattering ${\cal E} = 4$ and so precisely
one tree graph is possible for which $\sum_{id}'(d-2)V_{id} = 2$,
corresponding to $V_{44} = 1$ and all other $V_{id} = 0$. This
identifies the single graph which dominates the 4-point function
at low energies, and shows that the resulting leading energy
dependence is ${\cal A}_4(E) \sim E^4/(v^2 \,
m_{\scriptscriptstyle R}^2)$.

The utility of power-counting really becomes clear when subleading
behaviour is computed, so consider the size of the leading
corrections to the 4-point scattering amplitude. Order $E^6$
contributions are achieved if and only if either: ({\it i}) $L =
1$ and $V_{i4} = 1$, with all others zero; or ({\it ii}) $L = 0$
and $\sum_{i} \Bigl(4 V_{i6} + 2 V_{i4} \Bigr) = 4$. Since there
are no $d=2$ interactions, no one-loop graphs having 4 external
lines can be built using precisely one $d=4$ vertex and so only
tree graphs can contribute. Of these, the only two choices allowed
by ${\cal E} = 4$ at order $E^6$ are therefore the choices:
$V_{46} = 1$, or $V_{34} = 2$. Both of these contribute a result
of order ${\cal A}_4(E) \sim E^6/(v^2 \, m_{\scriptscriptstyle
R}^4)$.

\subsection{The Effective Lagrangian Logic}
With the power-counting results in hand we can see how to
calculate predictively --- {\it including loops} --- using the
non-renormalizable effective theory. The logic follows these
steps:

\begin{enumerate}
\item 
Choose the accuracy desired in the answer. (For instance an
accuracy of 1\% might be desired in a particular scattering
amplitude.)
\item 
Determine the order in the small ratio of scales ({\it i.e.}\ $r =
E/m_{\scriptscriptstyle R}$ in the toy model) which is required in
order to achieve the desired accuracy. (For instance if $r = 0.1$
then $O(r^2)$ is required to achieve 1\% accuracy.)
\item 
Use the powercounting results to identify which terms in ${\cal
L}_{\rm eff}$ can contribute to the observable of interest to the
desired order in $r$. At any fixed order in $r$ this always
requires a finite number (say: $N$) of terms in ${\cal L}_{\rm
eff}$ which can contribute.
\item[4a.] 
If the underlying theory is known, and is calculable, then compute
the required coefficients of the $N$ required effective
interactions to the accuracy required. (In the toy model this
corresponds to calculating the coefficients $a,b,c$ {\it etc.})
\item[4b.] 
If the underlying theory is unknown, or is too complicated to
permit the calculation of ${\cal L}_{\rm eff}$, then leave the $N$
required coefficients as free parameters. The procedure is
nevertheless predictive if more than $N$ observables can be
identified whose predictions depend only on these parameters.

\end{enumerate}

Step 4a is required when the low-energy expansion is being used as
an efficient means to accurately calculating observables in a
well-understood theory. It is the option of choosing instead Step
4b, however, which introduces much of the versatility of
effective-lagrangian methods. Step 4b is useful both when the
underlying theory is not known (such as when searching for physics
beyond the Standard Model) {\it and} when the underlying physics
is known but complicated (like when describing the low-energy
interactions of pions in Quantum Chromodynamics).

The effective lagrangian is in this way seen to be predictive even
though it is not renormalizable in the usual sense. In fact,
renormalizable theories are simply the special case of Step 4b
where one stops at order $r^0$, and so are the ones which dominate
in the limit that the light and heavy scales are very widely
separated. We see in this way {\it why} renormalizable
interactions play ubiquitous roles through physics! These
observations have important conceptual implications for the
quantum behaviour of other nonrenormalizable theories, such as
gravity, to which we return in the next section.

\subsubsection{The Choice of Variables}
The effective lagrangian of the toy model seems to carry much more
information when $\theta$ is used to represent the light particles
than it would if ${\cal I}$ were used. How can physics depend on
the fields which are used to parameterize the theory?

Physical quantities do not depend on what variables are used to
describe them, and the low-energy scattering amplitude is
suppressed by the same power of $r$ in the toy model regardless of
whether it is the effective lagrangian for ${\cal I}$ or $\theta$
which is used at an intermediate stage of the calculation.

The final result would nevertheless appear quite mysterious if
${\cal I}$ were used as the low-energy variable, since it would
emerge as a cancellation only at the end of the calculation. With
$\theta$ the result is instead manifest at every step. Although
the physics does not depend on the variables in terms of which it
is expressed, it nevertheless pays mortal physicists to use those
variables which make manifest the symmetries of the underlying
system.

\subsubsection{Regularization Dependence}
The definition of ${\cal L}_{\rm eff}$ appears to depend on lots
of calculational details, like the value of $\Lambda$ (or, in
dimensional regularization, the matching scale) and the minutae of
how the cutoff is implemented. Why doesn't ${\cal L}_{\rm eff}$
depend on all of these details?

${\cal L}_{\rm eff}$ generally {\sl does} depend on all of the
regularizational details. But these details all must cancel in
final expressions for physical quantities. Thus, some
$\Lambda$-dependence enters into scattering amplitudes through the
explicit dependence which is carried by the couplings of ${\cal
L}_{\rm eff}$ (beyond tree level). But $\Lambda$ also potentially
enters scattering amplitudes because loops over all light degrees
of freedom must be cut off at $\Lambda$ in the effective theory,
by definition. The cancellation of these two sources of
cutoff-dependence is guaranteed by the observation that $\Lambda$
enters only as a bookmark, keeping track of the light and heavy
degrees of freedom at intermediate steps of the calculation.

This cancellation of $\Lambda$ in all physical quantities ensures
that we are free to make any choice of cutoff which makes the
calculation convenient. After all, although all regularization
schemes for ${\cal L}_{\rm eff}$ give the same answers, more work
is required for some schemes than for others. Again, mere mortal
physicists use an inconvenient scheme at their own peril!

\medskip \noindent{\it Dimensional Regularization}

\medskip\noindent This freedom to use \emph{any} convenient scheme
is ultimately the reason why dimensional regularization may be
used when defining low-energy effective theories, even though the
dimensionally-regularized effective theories involve fields with
modes of arbitrarily high momentum. At first sight the necessity
of having modes of arbitrarily large momenta appear in
dimensionally-regularized theories would seem to make dimensional
regularization inconsistent with effective field theory. After
all, any dimensionally-regularized low-energy theory would appear
necessarily to include states having arbitrarily high energies.

In practice this is not a problem, so long as the effective
interactions are chosen to properly reproduce the
dimensionally-regularized scattering amplitudes of the full theory
(order-by-order in $1/M$). This is possible ultimately because the
difference between the cutoff- and dimensionally-regularized
low-energy theory can itself be parameterized by appropriate local
effective couplings within the low-energy theory. Consequently,
any regularization-dependent properties will necessarily drop out
of final physical results, once the (renormalized) effective
couplings are traded for physical observables.

In practice this means that one does not construct a
dimensionally-regularized effective theory by explicitly
performing a path integral over successively higher-energy
momentum modes of all fields in the underlying theory. Instead one
defines effective dimensionally regularized theories for which
heavy fields are completely removed. For instance, suppose it is
the low-energy influence of a heavy particle, $h$, having mass $M$
which is of interest. Then the high-energy theory consists of a
dimensionally-regularized collection of light fields, $\ell_i$,
and $h$, while the effective theory is a dimensionally-regularized
theory of the light fields, $\ell_i$, only. The effective
couplings of the low-energy theory are obtained by performing a
\emph{matching} calculation, whereby the couplings of the
low-energy effective theory are chosen to reproduce scattering
amplitudes or Green's functions of the underlying theory
order-by-order in powers of the inverse heavy scale, $1/M$. Once
the couplings of the effective theory are determined in this way
in terms of those of the underlying fundamental theory, they may
be used to compute any purely low-energy observable.

An important technical point arises if calculations are being done
to one-loop accuracy (or more) using dimensional regularization.
For these calculations it is convenient to trade the usual
minimal-subtraction (or modified-minimal-subtraction)
renormalization scheme, for a slightly modified \emph{decoupling
subtraction} scheme \cite{DSscheme}. In this scheme couplings are
defined using minimal (or modified-minimal) subtraction between
successive particle threshholds, with the couplings matched from
the underlying theory to the effective theory as each heavy
particle is successively integrated out. This results in a
renormalization group evolution of effective couplings which is
almost as simple as for minimal subtraction, but with the
advantage that the implications of heavy particles in running
couplings are explicitly decoupled as one passes to energies below
the heavy particle mass. See refs.~\cite{ETbooks, ETreviews} for
more details.

A great advantage of the dimensionally-regularized effective
theory is the absence of the cutoff scale $\Lambda$, which implies
that the only dimensionful scales which arise are physical
particle masses. This was implicitly used in the power-counting
arguments given earlier, wherein integrals over loop momenta were
replaced by powers of heavy masses on dimensional grounds. This
gives a sufficiently accurate estimate despite the ultraviolet
divergences in these integrals provided the integrals are
dimensionally regularized. For effective theories it is powers of
the arbitrary cutoff scale, $\Lambda$, which would arise in these
estimates, and because $\Lambda$ cancels out of physical
quantities, this just obscures how heavy physical masses appear in
the final results.

\subsection{The Meaning of Renormalizability}
The previous discussion about the cancellation between the cutoffs
on virtual light-particle momenta and the explicit
cutoff-dependence of ${\cal L}_{\rm eff}$ is eerily familiar. It
echoes the traditional discussion of the cancellation of the
regularized ultraviolet divergences of loop integrals against the
regularization dependence of the counterterms of the renormalized
lagrangian. There are, however, the following important
differences.

\begin{enumerate}
\item
The cancellations in the effective theory occur even though
$\Lambda$ is not sent to infinity, and even though ${\cal L}_{\rm
eff}$ contains arbitrarily many terms which are not renormalizable
in the traditional sense ({\it i.e.}\ terms whose coupling
constants have dimensions of inverse powers of mass in fundamental
units where $\hbar =  c = 1$).
\item
Whereas the cancellation of regularization dependence in the
traditional renormalization picture appears {\it ad-hoc} and
implausible, those in the effective lagrangian are sweet reason
personified. This is because they simply express the obvious fact
that $\Lambda$ only was introduced as an intermediate step in a
calculation, and so {\sl cannot} survive uncancelled in the
answer.
\end{enumerate}

This resemblance suggests Wilson's physical reinterpretation of
the renormalization procedure. Rather than considering a model's
classical lagrangian, such as ${\cal L}$ of
eq.~(\ref{abeltoymodel}), as something pristine and fundamental,
it is better to think of it also as an effective lagrangian
obtained by integrating out still more microscopic degrees of
freedom. The cancellation of the ultraviolet divergences in this
interpretation is simply the usual removal of an intermediate step
in an calculation to whose microscopic part we are not privy.



\section{Low-Energy Quantum Gravity}
\label{section:QGACP}

According to the approach just described, non-renormalizable
theories are not fundamentally different from renormalizable ones.
They simply differ in their sensitivity to more microscopic scales
which have been integrated out. It is instructive to see what this
implies for the non-renormalizable theories which sometimes are
required to successfully describe experiments. This is
particularly true for the most famous such case, Einstein's theory
of gravity. (See ref.~\cite{DonoghueRev} for another pedagogical
review of gravity as an effective theory.)

\subsection{General Relativity as an Effective Theory}

The low-energy degrees of freedom in this case are the metric,
$g_{\mu\nu}$, of spacetime itself. As has been seen in previous
sections, Einstein's action for this theory should be considered
to be just one term in a sum of all possible interactions which
are consistent with the symmetries of the low-energy theory (which
in this case are: general covariance and local Lorentz
invariance). Organizing the resulting action into powers of
derivatives of the metric leads to the following effective
lagrangian:
\begin{equation}
\label{gravaction}
 - \, {{\cal L}_{\rm eff} \over \sqrt{- g}} = \lambda
+ \frac{M_p^2}{2}  \, R + a \, R_{\mu\nu} \, R^{\mu\nu} + b \, R^2
+  d \, R_{\mu\nu\lambda\rho} R^{\mu\nu\lambda\rho} + e \, \Box R
+ {c \over m^2}\; R^3 + \cdots .
\end{equation}
Here $R_{\mu\nu\lambda\rho}$ is the metric's Riemann tensor,
$R_{\mu\nu}$ is its Ricci tensor, and $R$ is the Ricci scalar,
each of which involves precisely two derivatives of the metric.
For brevity only one representative example of the many possible
curvature-cubed terms is explicitly written. (We use here
Weinberg's curvature conventions \cite{GravCosm}, which differ
from those of Misner, Thorne and Wheeler \cite{MTW} by an overall
sign.)

The first term in eq.~(\ref{gravaction}) is the cosmological
constant, which is dropped in what follows since the observed size
of the universe implies $\lambda$ is extremely small. There is, of
course, no real theoretical understanding why the cosmological
constant should be this small. Once the cosmological term is
dropped, the leading term in the derivative expansion is the one
linear in $R$, which is the usual Einstein-Hilbert action of
General Relativity. Its coefficient defines Newton's constant (and
so also the Planck mass, $M_p^{-2} = 8 \pi G$).

The explicit mass scales, $m$ and $M_p$, are introduced to ensure
that the remaining constants $a,b,c,d$ and $e$ appearing in
eq.~(\ref{gravaction}) are dimensionless. Since it appears in the
denominator, the mass scale $m$ can be considered as the smallest
microscopic scale to have been integrated out to obtain
eq.~(\ref{gravaction}). For definiteness we might take the
electron mass, $m = 5\times 10^{-4}$ GeV, for $m$ when considering
applications at energies below the masses of all elementary
particles. (Notice that contributions like $m^2 R$ or $R^3/M_p^2$
could also exist, but these are completely negligible compared to
the terms displayed in eq.~(\ref{gravaction}).)

\subsubsection{Redundant Interactions}
As discussed in the previous section, some of the interactions in
the lagrangian (\ref{gravaction}) may be redundant, in the sense
that they do not contribute independently to physical observables
(like graviton scattering amplitudes about some fixed geometry,
say). To eliminate these we are free to drop any terms which are
either total derivatives or which vanish when evaluated at
solutions to the lower-order equations of motion.

The freedom to drop total derivatives allows us to set the
couplings $d$ and $e$ to zero. We can drop $e$ because $\sqrt{-g}
\, \Box R = \partial_\mu [\sqrt{-g} \, \nabla^\mu  R ]$, and we
can drop $d$ because the quantity
\begin{equation}
 X = R_{\mu\nu\lambda\rho} R^{\mu\nu\lambda\rho} -4 R_{\mu\nu}
 R^{\mu\nu} + R^2 \,,
\end{equation}
integrates to give a topological invariant in 4 dimensions. That
is, for a 4-dimensional manifold $\chi(M) = (1/32\pi^2)\int_M
\sqrt{g}\, X \, d^4x$ gives the Euler number of a compact manifold
$M$ -- and so $X$ is locally a total derivative. It is therefore
always possible to replace, for example, $R_{\mu\nu\lambda\rho}
R^{\mu\nu\lambda\rho}$ in the effective lagrangian with the linear
combination $4 \, R_{\mu\nu}R^{\mu\nu} - R^2$, with no
consequences for any observables which are insensitive to the
overall topology of spacetime (such as the classical equations, or
perturbative particle interactions). Any such observable therefore
is unchanged under the replacement $(a,b,d) \to (a',b',d') = (a -
d,b + 4d,0)$.

The freedom to perform field redefinitions allows further
simplification (just as was found for the toy model in earlier
sections). To see how this works, consider the infinitesimal field
redefinition $\delta g_{\mu\nu} = Y_{\mu\nu}$, under which the
leading term in ${\cal L}_{\rm eff}$ undergoes the variation
\begin{equation}
{M_p^2 \over 2} \; \delta \int d^4x \; \sqrt{-g} \, R =
 - \, {M_p^2 \over 2} \int d^4x \; \sqrt{-g} \left[ R^{\mu\nu} -
 \frac{1}{2} R g^{\mu\nu} \right] Y_{\mu\nu} \,.
\end{equation}
In particular, we may set the constants $a$ and $b$ to zero simply
by choosing $M_p^2 \, Y_{\mu\nu} = 2a\, R_{\mu\nu} - (a + 2b) \,
R\, g_{\mu\nu}$. Since the variation of the lower-order terms in
the action are always proportional to their equations of motion,
quite generally any term in ${\cal L}_{\rm eff}$ which vanishes on
use of the lower-order equations of motion can be removed in this
way (order by order in $1/m$ and $1/M_p$).

Since the lowest-order equations of motion or pure gravity
(without a cosmological constant) imply $R_{\mu\nu} = 0$, we see
that \emph{all} of the interactions beyond the Einstein-Hilbert
term which are explicitly written in eq.~(\ref{gravaction}) can be
removed in one of these two ways. The first interaction which can
have physical effects (for pure gravity with no cosmological
constant) in this low-energy expansion is therefore proportional
to the cube of the Riemann tensor.

This last conclusion changes if matter or a cosmological constant
are present, however, since then the lowest-order field equations
become $R_{\mu\nu} = S_{\mu\nu}$ for some nonzero tensor
$S_{\mu\nu}$. Then terms like $R^2$ or $R_{\mu\nu}R^{\mu\nu}$ no
longer vanish when evaluated at the solutions to the equations of
motion, but are instead equivalent to interactions of the form
$({S^\mu}_\mu)^2$, $S_{\mu\nu}R^{\mu\nu}$ or $S_{\mu\nu}
S^{\mu\nu}$. Since some of our later applications of ${\cal
L}_{\rm eff}$ are to the gravitational potential energy of various
localized energy sources, we shall find that these terms can
generate contact interactions amongst these sources.

\subsection{Power Counting}
Since gravitons are weakly coupled, perturbative power-counting
may be used to see how the high-energy scales $M_p$ and $m$ enter
into observables like graviton scattering amplitudes about some
fixed macroscopic metric. We now perform this power counting along
the lines of previous sections for the interactions of gravitons
near flat space: $g_{\mu\nu} = \eta_{\mu\nu} + h_{\mu\nu}$. For
the purposes of this power counting all we need to know about the
curvatures is that they each involve all possible powers of -- but
only two derivatives of -- the fluctuation field $h_{\mu\nu}$.

A power-counting estimate for the $L$-loop contribution to the
${\cal E}$-point graviton-scattering amplitude, ${\cal A}_{\cal
E}$, which involves $V_{id}$ vertices involving $d$ derivatives
and the emission or absorption of $i$ gravitons may be found by
arguments identical to those used previously for the toy model.
The main difference from the toy-model analysis is the existence
for gravity of interactions involving two derivatives, which all
come from the Einstein-Hilbert term in ${\cal L}_{\rm eff}$. (Such
terms also arise for Goldstone bosons for symmetry-breaking
patterns involving nonabelian groups and are easily incorporated
into the analysis.) The resulting estimate for ${\cal A}_{\cal E}$
turns out to be of order:
\begin{equation}
\label{GRcount1}
 {\cal A}_{\cal E}(E) \sim m^2 M_p^2 \left( {1
 \over M_p} \right)^{\cal E} \left( {m \over 4 \pi M_p} \right)^{2
 L} \left( {m^2 \over M_p^2} \right)^{Z} \left( {E \over m}
 \right)^P
\end{equation}
where $Z = \sum_{id}' V_{id}$ and $P = 2 + 2L + \sum_{id}' (d-2)
V_{id}$. The prime on both of these sums indicates the omission of
the case $d=2$ from the sum over $d$. If we instead group the
terms involving powers of $L$ and $V_{ik}$, eq.~(\ref{GRcount1})
takes the equivalent form
\begin{equation}
\label{GRcount1a}
 {\cal A}_{\cal E}(E) \sim E^2 M_p^2 \left( {1
 \over M_p} \right)^{\cal E}
 \left( {E \over 4 \pi M_p} \right)^{2
 L} {\prod_{i} \prod_{d>2}} \left[{E^2 \over M_p^2}
 \left( {E \over m} \right)^{(d-4)}  \right]^{V_{id}} \,.
\end{equation}
Notice that since $d$ is even, the condition $d > 2$ in the
product implies there are no negative powers of $E$ in this
expression.

Eqs.~(\ref{GRcount1}) and (\ref{GRcount1a}) share many of the
noteworthy features of eqs.~(\ref{PCresult1}) and
(\ref{PCresult2}). Again the weakness of the graviton's coupling
follows only from the low-energy approximations, $E \ll M_p$ and
$E \ll m$. When written as in eq.~(\ref{GRcount1a}), it is clear
that even though the ratio $E/m$ is potentially much larger than
$E/M_p$, it does not actually arise in ${\cal A}_{\cal E}$ unless
contributions from at least curvature-cubed interactions are
included (for which $d = 6$).

These expressions permit a determination of the dominant
low-energy contributions to scattering amplitudes. The minimum
suppression comes when $L = 0$ and $P = 2$, and so is given by
arbitrary tree graphs constructed purely from the Einstein-Hilbert
action. We are led in this way to what we are in any case inclined
to believe: it is classical General Relativity which governs the
low-energy dynamics of gravitational waves!

But the next-to-leading contributions are also quite interesting.
These arise in one of two ways, either: ({\it i}) $L = 1$ and
$V_{id} = 0$ for any $d\ne 2$; or ({\it ii}) $L = 0$, $\sum_i
V_{i4} = 1$, $V_{i2}$ is arbitrary, and all other $V_{id}$ vanish.
That is, the next to leading contribution is obtained by computing
the one-loop corrections using only Einstein gravity, or by
working to tree level and including precisely one
curvature-squared interaction in addition to any number of
interactions from the Einstein-Hilbert term. Both are suppressed
compared to the leading term by a factor of $(E/M_p)^2$, and the
one-loop contribution carries an additional factor of
$(1/4\pi)^2$.

For instance, for 2-body graviton scattering we have ${\cal E} =
4$, and so the above arguments imply the leading behaviour is
${\cal A}_4(E) \sim A_{2} (E/M_p)^2 + A_4 (E/M_p)^4 + \cdots$,
where the numbers $A_2$ and $A_4$ have been explicitly calculated.
At tree level all of the tree-level amplitudes turn out to vanish
except for those which are related by crossing symmetry to the
amplitude for which all graviton helicities have the same sign,
and this is given by \cite{DeWitt}:
\begin{equation}
 -i {\cal A}_{(++,++)}^{\rm tree} = 8 \pi  G \,\left(
 \frac{s^3}{tu} \right)\,,
\end{equation}
where $s$, $t$ and $u$ are the usual Mandelstam variables, all of
which are proportional to the square of the centre-of-mass energy,
$E_{\rm cm}$. Besides vindicating the power-counting expectation
that ${\cal A} \sim (E/M_p)^2$ to leading order, this example also
shows that the potentially frame-dependent energy, $E$, which is
relevant in the power-counting analysis is in this case really the
invariant centre-of-mass energy, $E_{\rm cm}$.

The one-loop correction to this result has also been computed
\cite{loopgraviton}, and is infrared divergent. These infrared
divergences cancel in the usual way with tree-level bremstrahlung
diagrams \cite{IRcancel}, leading to a finite result
\cite{DonoghueIR}, which is suppressed as expected relative to the
tree contribution by terms of order $(E/4 \pi M_p)^2$, up to
logarithmic corrections.

\subsubsection{Including Matter}
The observables of most practical interest for experimental
purposes involve the gravitational interactions of various kinds
of matter. It is therefore useful to generalize the previous
arguments to include matter and gravity coupled to one another. In
most situations this generalization is reasonably straightforward,
but somewhat paradoxically it is more difficult to treat the
interactions of non-relativistic matter than of relativistic
matter. This section describes the reasons for this difference.

\medskip\noindent {\it Relativistic Matter}

\medskip\noindent
Consider first relativistic matter coupled to gravity. Rather than
describing the general case, it suffices for the present purposes
to consider instead a relativistic boson (such as a massless
scalar or a photon) coupled to gravity but which does not
self-interact. The matter lagrangian for such a system is then
\begin{equation}
 - \, {{\cal L}_{\rm mat} \over \sqrt{-g}} = \frac{1}{2} \, g^{\mu\nu}
 \partial_\mu \phi \, \partial_\nu \phi + \frac{1}{4}
 F_{\mu\nu}F^{\mu\nu} \,,
\end{equation}
and so the new interaction terms all involve at most two matter
fields, two derivatives and any number of undifferentiated metric
fluctuations. This system is simple enough to include directly
into the above analysis provided the graviton-matter vertices are
counted together with those from the Einstein-Hilbert term when
counting the vertices having precisely two derivatives with
$V_{i2}$.

Particular kinds of higher-derivative terms involving the matter
fields may also be included equally trivially, provided the mass
scales which appear in these terms appear in the same way as they
did for the graviton. For instance, scalar functions built from
arbitrary powers of $A_{\mu}/M_p$ and its derivatives
$\partial_\mu/m$ can be included, along the lines of
\begin{equation}
 \Delta {\cal L}_n = c_{kn}\, m^2 M_p^2 \, \left(
 \frac{{F}_{\mu\nu} \Box^k {F}^{\mu\nu}}{m^{2+2k} M_p^{2}} \right)^n
 \,.
\end{equation}
If the dimensionless constant $c_{kn}$ in these expressions is
$O(1)$, then the power-counting result for the dependence of
amplitudes on $m$ and $M_p$ is the same as it is for the
pure-gravity theory, with vertices formed from $\Delta {\cal L}_n$
counted amongst those with $d = (2 + 2k) n$ derivatives. If $m \ll
M_p$ it is more likely that powers of $A_\mu$ come suppressed by
inverse powers of $m$ rather than $M_p$, however, in which case
additional $A_\mu$ vertices are less suppressed than would be
indicated above. The extension of the earlier power-counting
estimate to this more general situation is straightforward.

Similar estimates also apply if a mass, $m_\phi$, for the scalar
field is included, provided that this mass is not larger than the
energy flowing through the external lines: $m_\phi \lsim E$. This
kind of mass does not change the power-counting result appreciably
for observables which are infrared finite (which may require, as
mentioned above) summing over an indeterminate number of soft
final gravitons. They do not change the result because
infrared-finite quantities are at most logarithmically singular as
$m_\phi \to 0$ \cite{RGgoodthing}, and so their expansion in
$m_\phi / E$ simply adds terms for which factors of $E$ are
replaced by smaller factors of $m_\phi$. But the above discussion
can change dramatically if $m_\phi \gg E$, since an important
ingredient in the dimensional estimate is the assumption that the
largest scale in the graph is the external energy $E$.
Consequently the power-counting given above only directly applies
to relativistic particles.

\medskip\noindent {\it Non-relativistic Matter}

\medskip\noindent
The situation is more complicated if the matter particles move
non-relativistically since in this case the particle mass is much
larger than the momenta involved in the external lines, $p = |{\bf
p}| \ll m_\phi$, so $E \approx m_\phi + p^2/(2m_\phi) + \cdots$.
We expect quantum corrections to the gravitational interactions of
such particles also to be suppressed (such as, for instance, for
the Earth) despite the energies and momenta involved being much
\emph{larger} than $M_p$. Indeed, most of the tests of general
relativity involve the gravitational interactions and orbits of
very non-relativistic `particles', like planets and stars. How can
this be understood?

The case of non-relativistic particles is also of real practical
interest for the applications of effective field theories in other
branches of physics. This is so, even though one might think that
an effective theory should contain only particles which are very
light. Non-relativistic particles can nevertheless arise in
practice within an effective field theory, even particles having
masses which are large compared to those of the particles which
were integrated out to produce the effective field theory in the
first place. Such massive particles can appear consistently in a
low-energy theory provided they are stable (or extremely
long-lived) and so cannot decay and release enough energy to
invalidate the low-energy approximation. Some well-known examples
of this include the low-energy nuclear interactions of nucleons
(as described within chiral perturbation theory \cite{NchiPT}),
the interactions of heavy fermions like the $b$ and $t$ quark (as
described by heavy-quark effective theory (HQET) \cite{HQET}) and
the interactions of electrons and nuclei in atomic physics (as
described by non-relativistic Quantum Electrodynamics (NRQED)
\cite{NRQED}).

The key to understanding the effective field theory for very
massive, stable particles at low energies lies in the recognition
that their anti-particles need \emph{not} be included since they
would have already been integrated out to obtain the effective
field theory of interest. As a result heavy-particle lines within
the Feynman graphs of the effective theory only directly connect
external lines, and never arise as closed loops.

The most direct approach to the estimating the size of quantum
corrections in this case is to power-count as before, subject to
the restriction that graphs including internal closed loops of
heavy particles are to be excluded. Ref.~\cite{DonoghuePC} has
performed such a power-counting analysis along these lines, and
shows that quantum effects remain suppressed by powers of
\emph{light}-particle energies (or small momentum transfers)
divided by $M_p$ through the first few nontrivial orders of
perturbation theory. Although heavy-particle momenta can be large,
$p \gg M_p$, they only arise in physical quantities through the
small relativistic parameter, $p/m_\phi \sim v$, rather than
through $p/M_p$, extending the suppression of quantum effects
obtained earlier to non-relativistic problems.

Unfortunately, if a calculation is performed within a covariant
gauge, individual Feynman graphs \emph{can} depend on large powers
like $m_\phi/M_p$, even though these all cancel in physical
amplitudes. For this reason an all-orders, inductive, proof of the
above power-counting remains elusive. As ref.~\cite{DonoghuePC}
also makes clear, progress towards such an all-orders
power-counting result is likely to be easiest within a physical,
non-covariant gauge, since such a gauge allows powers of small
quantities like $v$ to be most easily followed.

\medskip\noindent{\it Non-relativistic Effective Field Theory}

\medskip\noindent
If experience with electromagnetism is any guide, effective field
theory techniques are also likely to be the most efficient way to
systematically keep track of both the expansion in inverse powers
of both heavy masses, $1/M_p$ and $1/m_\phi$ --- particularly for
bound orbits. Relative to the theories considered to this point,
the effective field theory of interest has two unusual properties.
First, since it involves very slowly-moving particles, Lorentz
invariance is not simply realized on the corresponding
heavy-particle fields. Second, since the effective theory does not
contain antiparticles for the heavy particles, the heavy fields
which describe them contain only positive-frequency parts. To
illustrate how these features arise, we briefly sketch how such a
non-relativistic effective theory arises once the antiparticles
corresponding to a heavy stable particle are integrated out. We do
so using a toy model of a single massive scalar field, and we work
in position space to facilitate the identification of the
couplings to gravitational fields.

Consider, then, a complex massive scalar field (we take a complex
field to ensure low-energy conservation of heavy-particle number)
having action
\begin{equation} \label{PhiAction}
 - \frac{{\cal L}}{\sqrt{-g}} = g^{\mu\nu} \partial_\mu
 \phi^* \partial_\mu \phi
 + m_\phi^2 \phi^*\phi
 \, ,
\end{equation}
for which the conserved current for heavy-particle number is
\begin{equation}
 J^\mu = -i g^{\mu\nu} (\phi^* \partial_\nu \phi -
 \partial_\nu \phi^* \phi)  \, .
\end{equation}
Our interest is in exhibiting the leading couplings of this field
to gravity, organized in inverse powers of $m_\phi$. We imagine
therefore a family of observers relative to whom the heavy
particles move non-relativistically, and whose foliation of
spacetime allows the metric to be written
\begin{equation} \label{foliation}
 ds^2 = - (1 + 2\varphi) \, dt^2 + 2 N_i \, dt \, dx^i + \gamma_{ij} \,
 dx^i \, dx^j \, ,
\end{equation}
where $i = 1,2,3$ labels coordinates along the spatial slices
which these observers define.

When treating non-relativistic particles it is convenient to
remove the rest mass of the heavy particle from the energy, since
(by assumption) this energy is not available to other particles in
the low-energy theory. For the observers just described this can
be done by extracting a time-dependent phase from the
heavy-particle field according to $\phi(x) = F e^{-i m_\phi t} \,
\chi(x)$. $F = (2 m_\phi)^{-1/2}$ is chosen for later convenience,
to ensure a conventional normalization for the field $\chi$. With
this choice we have $\partial_t \phi = F (\partial_t - i m_\phi)
\chi$, and the extra $m_\phi$ dependence introduced this way has
the effect of making the large-$m_\phi$ limit of the
positive-frequency part of a relativistic action easier to follow.

With these variables the action for the scalar field becomes
\begin{equation} \label{NRLagr}
 - \frac{{\cal L}}{\sqrt{-g}}
 = \frac{m_\phi}{2} \left( g^{tt} + 1 \right) \chi^* \chi +
 \frac{i}{2} \, g^{t\mu} \left(\chi^* \partial_\mu \chi -
 \partial_\mu \chi^* \chi \right) + \frac{1}{2 m_\phi} \,
 g^{\mu\nu} \,
 \partial_\mu \chi^* \partial_\nu \chi \,,
\end{equation}
and the conserved current for heavy-particle number becomes
\begin{equation} \label{NRcurrent}
 J^\mu = - g^{\mu t} \, \chi^*\chi - \frac{i}{2m_\phi} \, g^{\mu\nu}
 \, ( \chi^* \partial_\nu \chi - \partial_\nu \chi^* \chi) \, .
\end{equation}
Here $g^{tt} = -1/D$, $g^{ti} = N^i/D$ and $g^{ij} = \gamma^{ij} -
N^i N^j/D$, with $N^i = \gamma^{ij}N_j$ and $D = 1 + 2\phi +
\gamma^{ij}N_i N_j$.

Notice that for Minkowski space, where $g^{\mu\nu} = \eta^{\mu\nu}
= {\rm diag}(-+++)$, the first term in ${\cal L}$ vanishes,
leaving a result which is finite in the $m_\phi \to \infty$ limit.
Furthermore --- keeping in mind that the leading time and space
derivatives are of the same order of magnitude ($\partial_t \sim
\nabla^2/m_\phi$) --- the leading large-$m_\phi$ part of ${\cal
L}$ is equivalent to the usual non-relativistic Schr\"odinger
lagrangian density, ${\cal L}_{\rm sch} = \chi^* \left[
i\partial_t + \nabla^2 /(2m_\phi)\right] \chi $. In the same limit
the density of $\chi$ particles also takes the standard
Schr\"odinger form $\rho = J^t = \chi^* \chi + O(1/m_\phi)$.

The next step consists of integrating out the anti-particles,
which (by assumption) cannot be produced by the low-energy physics
of interest. In principle, this can be done by splitting the
relativistic field, $\chi$, into its positive- and
negative-frequency parts, $\chi_{(\pm)}$, and performing the
functional integral over the negative-frequency part,
$\chi_{(-)}$. (To leading order this often simply corresponds to
setting the negative-frequency part to zero.) Once this has been
done the fields describing the heavy particles have the
non-relativistic expansion
\begin{equation}
 \chi_{(+)}(x) = \int d^3p \; a_p \, u^{(+)}_p(x) \,,
\end{equation}
with no anti-particle term involving $a^*_p$. It is this step
which ensures the absence of virtual heavy-particle loops in the
graphical expansion of amplitudes in the low-energy effective
theory.

Writing the heavy-particle action in this way extends the standard
parameterized post-Newtonian (PPN) expansion \cite{PPN},
\cite{GravCosm} to the effective quantum theory, and so forms the
natural setting for an all-orders power-counting analysis which
keeps track of both quantum and relativistic effects. For
instance, for weak gravitational fields having $\phi \sim N^2 \sim
\gamma_{ij} - \delta_{ij} \ll 1$, the leading gravitational
coupling for large $m_\phi$ may be read off from
eq.~(\ref{NRLagr}) to be
\begin{equation}
 {\cal L}_0 \approx - \frac{m_\phi}{2} \sqrt{-g} \left( g^{tt} + 1 \right)
 \chi^* \chi  \approx - m_\phi \left(\phi + \frac{N^2}{2} \right)
 \chi^* \chi \,,
\end{equation}
which for $N_i = 0$ reproduces the usual Newtonian coupling of the
potential $\phi$ to the nonrelativistic mass distribution. For
several $\chi$ particles prepared in position eigenstates we are
led in this way to considering the gravitational field of a
collection of classical point sources.

The real power of the effective theory lies in identifying the
subdominant contributions in powers of $1/m_\phi$, however, and
the above discussion shows that different components of the metric
couple to matter at different orders in this small quantity. Once
$\phi$ is shifted by the static non-relativistic Newtonian
potential, however, the remaining contributions are seen to couple
with a strength which is suppressed by negative powers of
$m_\phi$, rather than positive powers. A full power-counting
analysis using such an effective theory, along the lines of the
analogous electromagnetic problems \cite{NRQED}, would be very
instructive.

\subsection{Effective Field Theory in Curved Space}
\label{section:QFTCurved}
All of the explicit calculations of the previous sections are
performed for weak gravitational fields, which are well described
as perturbations about flat space. This has the great virtue of
being sufficiently simple to make explicit calculations possible,
including the widespread use of momentum-space techniques. Much
less is known in detail about effective field theory in more
general curved spaces, although its validity is implicitly assumed
by the many extant calculation of quantum effects in curved space
\cite{BirrellDavies}, including the famous calculation of Hawking
radiation \cite{Hawking} by black holes. This section provides a
brief sketch of some effective-field theory issues which arise in
curved-space applications.

For quantum systems in curved space we are still interested in the
gravitational action, eq.~(\ref{gravaction}), possibly
supplemented by a matter action such as that of
eq.~(\ref{PhiAction}). As before, quantization is performed by
splitting the metric into a classical background,
$\hat{g}_{\mu\nu}$, and a quantum fluctuation, $h_{\mu\nu}$,
according to $g_{\mu\nu} = \hat{g}_{\mu\nu} + h_{\mu\nu}$. A
similar expansion may be required for the matter fields, $\phi =
\hat{\phi} + \varphi$, if these acquire vacuum expectation values,
$\hat{\phi}$.

The main difference from previous sections is that
$\hat{g}_{\mu\nu}$ is not assumed to be the Minkowski metric.
Typically we imagine the spacetime may be foliated by a set of
observers, as in eq.~(\ref{foliation}),
\begin{equation} \label{foliation1}
 ds^2 = - (1 + 2\varphi) \, dt^2 + 2 N_i \, dt \, dx^i
 + \gamma_{ij} \,
 dx^i \, dx^j \, .
\end{equation}
and we imagine that the slices of constant $t$ are Cauchy
surfaces, in the sense that the future evolution of the fields is
uniquely specified by initial data on a constant-$t$ slice. The
validity of this initial-value problem may also require boundary
conditions for the fluctuations, $h_{\mu\nu}$, such as that they
vanish at spatial infinity.

\subsubsection{When Should Effective Lagrangians Work?}
Many of the general issues which arise in this problem are similar
to those which arise outside of gravitational physics when quantum
fields are considered in the presence of background classical
scalar or electromagnetic fields. In particular, there is a
qualitative difference between the cases where the background
fields are time-independent or time-dependent. The purpose of this
section is to argue that we expect an effective field theory to
work provided that the background fields vary sufficiently slowly
with respect to time, an argument which in its relativistic
context is called the `nice slice' criterion \cite{NiceSlice}.


\medskip\noindent{\it Time-Independent Backgrounds}

\medskip\noindent
In flat space, background fields which are time-translation
invariant allow the construction of a conserved energy, $H$, which
evolves forward the fields from one $t$ slice to another. If $H$
is bounded below then a stable ground state for the quantum fields
(the `vacuum') typically exists, about which small fluctuations
can be explored perturbatively. (Examples where $H$ is \emph{not}
bounded from below can arise, such as for charged particles in a
sufficiently strong background electric field \cite{Efield}. In
such situations the runaway descent of the system to arbitrarily
low energies is interpreted as being due to continual particle
pair production by the background field.) Since energy can be
defined and is conserved, it can be used as a criterion for
defining an effective theory which distinguishes between states
which are `low-energy' or `high-energy' as measured using $H$.
Once this is possible, a low-energy effective theory may be
defined, and we expect the general uncertainty-principle arguments
given earlier to ensure that it is local in time.

A similar statement holds for background gravitational fields,
with a conserved Hamiltonian following from the existence of a
time-like Killing vector field, $K^\mu$, for the background metric
$\hat{g}_{\mu\nu}$. (A Killing vector field satisfies the
condition $\hat\nabla_\mu \hat{K}_\nu + \hat \nabla_\nu
\hat{K}_\mu = 0$, which is the coordinate-invariant expression of
the existence of a time-translation invariance of the background
metric. The carets indicate that the derivatives are defined, and
the indices are lowered, by the background metric
$\hat{g}_{\mu\nu}$.) For matter fields $H$ is defined in terms of
$K^\mu$ and the stress tensor $T_{\mu\nu}$ by the integral
\begin{equation} \label{Hmatdef}
 H_{\rm mat} = \int_t d\Sigma_\mu \, K^\nu \, {T^\mu}_\nu \,.
\end{equation}
Here the integral is over any spacelike slice whose measure is
$d\Sigma_\mu$. Given appropriate boundary conditions this
definition can also be extended to the fluctuations, $h_{\mu\nu}$,
of the gravitational field \cite{CanonQG}, \cite{Hdef}.

If the timelike Killing vector, $K^\mu$, is hypersurface
orthogonal, then the spacetime is called static, and it is
possible to adapt our coordinates so that $K^\mu \partial_\mu =
\partial_t$ and so the metric of eq.~(\ref{foliation1})
specializes to $N_i = 0$, with $\phi$ and $\gamma_{ij}$
independent of $t$. In this case we call the observers
corresponding to these coordinates `Killing' observers. In order
for these observers to describe physics everywhere, it is implicit
that the timelike Killing vector $K^\mu$ be globally defined
throughout the spacetime of interest.

If $H$ is defined and is bounded from below, its lowest eigenstate
defines a stable vacuum and allows the creation of a Fock space of
fluctuations about this vacuum.  As was true for non-gravitational
background fields, in such a case we might again expect to be able
to define an effective theory, using $H$ to distinguish
`low-energy' from `high-energy' fluctuations about any given
vacuum.

It is often true that $K^\mu$ is not unique because there is more
than one globally-defined timelike Killing vector in a given
spacetime. For instance, this occurs in flat space where different
inertial observers can be rotated, translated or boosted relative
to one another. In this case the Fock space of states to which
each of the observers is led are typically all unitarily
equivalent to one another, and so each observer has an equivalent
description of the physics of the system.

An important complication to this picture arises when the timelike
Killing vector field cannot be globally defined throughout all of
the spacetime. In this case horizons exist, which divide the
regions having timelike Killing vectors from those which do not.
Often the Killing vector of interest becomes null at the
boundaries of these regions. Examples of this type include the
accelerated `Rindler' observers of flat space, as well as the
static observers outside of a black hole. In this case it is
impossible to foliate the entire spacetime using static slices
which are adapted to the Killing observer, and the above
construction must be reconsidered.

Putting the case of horizons aside for the moment, we expect that
a sensible low-energy/high-energy split should be possible if the
background spacetime is everywhere static, and if the conserved
energy $H$ is bounded from below.


\medskip\noindent{\it Time-Dependent Backgrounds}

\medskip\noindent
In non-gravitational problems, time-dependence of the background
fields need not completely destroy the utility of a Hamiltonian or
of a ground state provided that this time dependence is
sufficiently weak as to be treated adiabatically. In this case the
lowest eigenstate of $H(t)$ for any given time defines both an
approximate ground state and an energy in terms of which
low-energy and high-energy states can be defined (such as by using
an appropriate eigenvalue, $\Lambda(t)$, of $H(t)$).

Once the system is partitioned in this way into low-energy and
high-energy state, one can ask whether a purely low-energy
description of time evolution is possible using only a low-energy,
local effective lagrangian. The main danger which arises with
time-dependent backgrounds is that the time evolution of the
system need not keep low-energy states at low energies, or
high-energy states at high energies. There are several ways in
which this might happen.
\begin{itemize}
\item
The time dependence of the background can be rapid, for instance
having Fourier components which are larger than the dividing line,
$\Lambda(t)$, between low- and high-energy states. If so, the
background evolution is not sufficiently adiabatic to prevent the
direct production of `heavy' particles, and an effective field
theory need not exist. The criterion for this not to happen would
be that the energies of any produced particles be smaller than the
dividing eigenvalue, $\Lambda(t)$, which typically requires that
$\Lambda(t)$, be much smaller than the background time dependence,
for instance if: $\Lambda \ll \dot\Lambda/\Lambda$.
\item
Even for slowly-varying backgrounds there can be other dangers,
such as the danger of level-crossing. For instance the dividing
eigenvalue, $\Lambda(t)$, may slowly evolve to smaller values and
so eventually invalidate an expansion in inverse powers of
$\Lambda(t)$. In this case states which are prepared in the
low-energy regime may evolve out of it, leading to a breakdown in
the low-energy approximation. For instance, this might happen if
$\Lambda(t)$ were chosen to be the mass of an initially heavy
particle, which is integrated out to obtain the effective theory.
If evolution of the background fields allows this mass to evolve
to become too low, then eventually it becomes a bad approximation
to have integrated it out.
\item
A related potential problem can arise if states whose energy is
initially larger than $\Lambda(t)$ enter the low-energy theory by
evolving into states having energy lower than $\Lambda(t)$. (For
example, this could happen for a charged particle in a decreasing
magnetic field if the effective theory is set up so that the
dividing energy, $\Lambda(t)$, is not time dependent. Then Landau
levels continuously enter the low-energy theory as the magnetic
field strength wanes.) This usually is only a problem for the
effective-theory formulation if the states which enter in this way
are not in their ground state when they do so, since then new
physical excitations would arise at low energies. Conversely, they
are not a problem for the effective field theory description so
long as they enter in their ground states, as can usually be
ensured if the background time evolution is adiabatic.
\end{itemize}
What emerges from this is that an effective field theory can make
sense despite the presence of time-dependent backgrounds, provided
one can focus on the evolution of low-energy states, ($E<
\Lambda(t)$), without worrying about losing probability into
high-energy states ($E > \Lambda(t)$). This is usually ensured if
the background time evolution is sufficiently adiabatic.

A similar story should also hold for background spacetimes which
are not globally static, but for which a globally-defined timelike
hypersurface-orthogonal vector field, $V^\mu$, exists. For such a
spacetime the observers for whom $V^\mu$ is tangent to world-lines
can define a foliation of spacetime, as in eq.~(\ref{foliation1}),
but with the various metric components not being $t$-independent.
In this case the quantity defined by eq.~(\ref{Hmatdef}) need not
be conserved, $H = H(t)$, for these observers. A low-energy
effective theory should nonetheless be possible, provided $H(t)$
is bounded from below and is sufficiently slowly varying (in the
senses described above). If such a foliation of spacetime exists,
following ref.~\cite{NiceSlice} we call it a `nice slice'.

\subsubsection{General Power Counting}
Given an effective field theory, the next question is to analyze
systematically how small energy ratios arise within perturbation
theory. Since the key power-counting arguments of the previous
sections were given in momentum space, a natural question is to
ask how much of the previous discussion need apply to quantum
fluctuations about more general curved spaces. In particular, does
the argument that shows how large quantum effects are at
arbitrary-loop order apply more generally to quantum field theory
in curved space?

An estimate of higher-loop contributions performed in position
space is required in order to properly apply the previous
arguments to more general settings. Such a calculation is possible
because the crucial part of the earlier estimate relied on an
estimate for the high-energy dependence of a generic Feynman
graph. This estimate was possible on dimensional grounds given the
high-energy behaviour of the relevant vertices and propagators.
The analogous computation in position space is also possible,
where it instead relies on the \emph{operator product expansion}
\cite{OPE}. In position space one's interest is in how effective
amplitudes behave in the short-distance regime, rather than the
limit of high energy. But the short-distance limit of propagators
and vertices are equally well-known, and resemble the
short-distance limit which is obtained on flat space. Consequently
general statements can be made about the contributions to the
low-energy effective theory.

Physically, the equivalence of the short-distance position-space
and high-energy momentum-space estimates is expected because the
high-energy contributions arise due to the propagation of modes
having very small wavelength, $\lambda$. Provided this wavelength
is very small compared with the local radius of curvature, $r_c$,
particle propagation should behave just as if it had taken place
in flat space. One expects the most singular behaviour to be just
as for flat space, with curvature effects appearing in subdominant
corrections as powers of $\lambda/r_c$.

Unfortunately, although the result is not in serious doubt, such a
general position-space estimates for gravitational physics on
curved space has yet been done explicitly at arbitrary orders of
perturbation theory. Partial results are known, however, including
general calculations of the leading one-loop ultraviolet
divergences in curved space \cite{Gilkey}.

\subsubsection{Horizons and Large Redshifts}
Among the most interesting applications of effective field theory
ideas to curved space is the study of quantum effects near black
holes and in the early universe. In particular, for massive black
holes ($M \gg M_p$) one expects semi-classical arguments to be
valid since the curvature at the horizon is small and the
interesting phenomena (like Hawking radiation) rely only on the
existence of the horizon rather than on any properties of the
spacetime near the internal curvature singularity
\cite{LivRev-BHT}. Although the power-counting near the horizon
has not been done in the same detail as it has been for the
asymptotic regions, semi-classical effective-field-theory
arguments at the horizon are expected to be valid. Similar
statements are also expected to be true for calculations of
particle production in inflationary universes.

An objection has been raised to the validity of effective field
theory arguments in both the black hole \cite{Unruh}, \cite{Ted}
and inflationary \cite{TedReview}, \cite{Robert} contexts. For
both of these cases the potential difficulty arises if one
compares the energy of the modes as measured by different
observers situated throughout the spacetime. For instance, a mode
which emerges far from a black hole at late times with an energy
(as seen by static and freely-falling observers) close to the
Hawking temperature starts off having extremely high energies as
seen by freely-falling observers very close to, but outside of,
the black hole's event horizon just as it forms. The energy
measured at infinity is much smaller because the state experiences
an extremely large redshift as it climbs out of the black hole's
gravitational well. The corresponding situation in inflation is
the phenomenon in which modes get enormously redshifted (all the
way from microscopic to cosmological scales) as the universe
expands.

It has been argued that these effects prevent a consistent
low-energy effective theory from being built in these situations,
because very high-energy states are continuously turning up at
later times at low energies. If so, this would seem to imply that
a reliable calculation of phenomena like Hawking radiation (or
inflationary fluctuations of the CMB) necessarily require an
understanding of very high energy physics. Since we do not know
what this very high energy physics is, this is another way of
saying that these predictions are theoretically unreliable, since
uncontrolled theoretical errors potentially contribute with the
same size as the predicted effect.

The remainder of this section argues that although the concerns
raised are legitimate, they are special cases of the general
conditions mentioned earlier which govern the applicability of
effective field theory ideas to time-dependent backgrounds also in
non-gravitational settings. As such one expects to find robustness
against adiabatic physics at high energies, and sensitivity to
non-adiabatic effects. (This expectation is borne out by the explicit
calculations to date.) Given a concrete theory of what the
high-energy physics is, one can then ask into which category it
falls, and so better quantify the theoretical error.

\medskip\noindent{\it Black Holes}

\medskip\noindent
For black holes the potential high-energy difficulty can be traced
to the fact that freely-falling static observer are strongly boosted relative
to one another near the
event horizon. This is why the escaping modes have such high
energies as seen by freely-falling observers near the horizon. This raises
two separate issues for effective field theories, which are worth
separating: the issue of the relevance of freely-falling observers
measuring very large energies for escaping objects, and the
issue of high-energy states descending into the low-energy theory
as time progresses. Each of these is now discussed in turn.

The fact that freely-falling observers measure different energies
for outgoing particles, depending on their distance from the
horizon, underlines that there is a certain amount of frame
dependence in any effective-theory description, even in flat
space. This is so because energy is used as the criterion for
deciding which states fall into the effective theory and which do
not, yet any nominally low-energy particle has a large energy as
seen by a sufficiently boosted observer. In practice this is not a
problem because the validity of the effective theory description
only requires the \emph{existence} of low-energy observers, not
that all observers be at low energy. What is important is that the
physically relevant energies for the process of interest --- for
instance, the centre-of-mass energies in a scattering event ---
are small in order for this process to be describable using a
low-energy theory. Once this is true, invariant quantities like
cross sections take a simple low-energy form when expressed in
terms of physical kinematic variables, regardless of the energies
which the particles involved have as measured by observers who are
highly boosted compared to the centre-of-mass frame.

Flat-space experience therefore suggests that there need not be a
problem associated with escaping modes having large energies as
seen by freely-falling observers. This only indicates that the use
of some observers near the horizon may be problematic. So long as
the physics involved does not rely crucially on these observers,
it may in any case allow an effective-theory description. This is
essentially the point of view put forward in refs.~\cite{Ted} and
\cite{NiceSlice},\footnote{These authors have slightly different
spins on the more philosophical question of whether
trans-Planckian physics is likely to be found to be
non-adiabatic.} which argue that the robustness of the Hawking
radiation to high-energy physics is most simply understood if one
is careful to foliate the spacetime using slices which are chosen
to be `nice slices', in the sense described above, which cut
through the horizon in such a way as to encounter only small
curvatures and adiabatic time variation. Since such slices exist,
a low-energy theory may be set up in terms of the slowly-varying
$H(t)$ which these slices define \cite{NSHawking}.

Of course, calculations need not explicitly use the nice slices in
order to profit from their existence. In the same way that
dimensional regularization can be more useful in practice for
calculations in effective field theories, despite its inclusion of
arbitrarily high energy modes, the sensitivity of Hawking
radiation to high energies can be investigated using a convenient
covariant regularization. This is because if nice slices exist,
covariant calculations must reproduce the insensitivity to high
energies which they guarantee. This is borne out by explicit
calculations of the sensitivity of the Hawking radiation to high
energies \cite{Hambli} using a simple covariant regularization.

The most important manner in which high-energy states can
influence the Hawking radiation has been identified from
non-covariant studies, such as those which model the high-energy
physics as non-Lorentz invariant dispersion relations for
otherwise free particles \cite{DispRel}. (See
ref.~\cite{TedReview} for a review, with references, of these
calculations.) These identify the second pertinent issue mentioned
above: the descent of higher-energy states into the low-energy
theory. In these calculations high-energy modes cross into the
low-energy theory because of their redshift as they climb out of
the gravitational potential well of the black hole. The usual
expression for the Hawking radiation follows provided that these
modes enter the low-energy theory near the black hole horizon in
their adiabatic ground state (a result which can also be seen in
covariant approaches, where it can be shown that the Hawking
radiation depends only on the form of the singularity of the
propagator near the light cone \cite{ShortDist}). If these modes
do not start off in their ground state, then they potentially
cause observable changes to the Hawking radiation.

The condition that high-energy modes enter the low-energy theory
in their ground state is reminiscent of the same condition which
was encountered in previous sections as a general pre-condition
for the validity of a low-energy effective description when there
are time-dependent backgrounds (including, for example, the
descent of Landau levels in a decreasing magnetic field). In
non-gravitational contexts it is automatically satisfied if the
background evolution is adiabatic, and this can also be expected
to be true in the gravitational case. Of course, this expectation
cannot be checked explicitly unless the theory for the relevant
high-energy physics is specified, but it is borne out by all of
the existing calculations. To the extent that high-energy modes do
not arise in their adiabatic vacua, their effects might be
observable in the Hawking radiation as well as in possibly many
other observables which would otherwise be expected to be
insensitive to high-energy physics.

Clearly this is good news, since it tells us that we can believe
that generic quantum effects do not ruin the classical
calculations using General Relativity which tell us that black
holes exist. Nor do they ruin the semiclassical calculations which
lead to effects like the Hawking radiation \cite{Hawking} in the
vicinity of black holes --- provided that the black hole mass is
much larger than $M_p$ (which we shall see is required if quantum
effects are to remains small at the event horizon). On the other
hand, it means that we cannot predict the final stages of
black-hole evaporation, since these inevitably lead to small black
hole masses, where the semiclassical approximation breaks down.

\medskip\noindent{\it Inflation}

\medskip\noindent
Many of the issues concerning the validity of effective field
theories which arise for the Hawking radiation also arise within
inflationary cosmology, and has generated considerable discussion
due to the recent advent of precise measurements of CMB
temperature fluctuations. By analogy with the black hole case, it
has been proposed \cite{TedReview}, \cite{Robert} that
very-high-energy physics may not decouple from inflationary
predictions due to the exponential expansion of space. If so,
there might be detectable imprints on the observed temperature
fluctuations in the cosmic microwave background \cite{LivRev-CMB}.
Conversely, if high-energy effects \emph{do} contaminate
inflationary predictions for CMB fluctuations at an observable
level then inflationary predictions themselves must be recognized
as containing an uncontrollable theoretical uncertainty. If so,
their successful description of the observations cannot be deemed
to be credible evidence of the existence of an earlier
inflationary phase. There is clearly much at stake.

It is beyond the scope of the this article to summarize all of the
intricacies associated with quantum field theory in de Sitter
space, so we focus only on the parallels with the black hole
situation. The bottom line for cosmology is similar to what was
found for the Hawking radiation.
\begin{itemize}
\item
Observable effects \emph{can} be obtained from physics associated
with energies, $E$, much higher than the inflationary expansion
rate, $H$, if the states associated with the heavy physics are
chosen \emph{not} to be in their adiabatic vacuum. (In the
inflationary context we take `adiabatic vacuum' to mean the
Bunch-Davies vacuum \cite{BD}. See, however,
refs.~\cite{BDrequired} for arguments against the use of
non-standard vacua in de Sitter space.) Potentially observable
effects have been obtained by explicit calculations which
incorporate non-adiabatic physics of various types. For many of
these the high-energy physics is modelled as a free particle
having a non-Lorentz-invariant dispersion relation
\cite{RobertReview}. However large Lorentz-violating interactions
need not be required since similar effects are also obtained using
more conventional inflationary field theories, for which
background scalar fields are allowed to roll non-adiabatically
\cite{TPI1}.
\item
If the states associated with high-energy physics are prepared in
their adiabatic vacua then an effective field theory description
applies. In this case most kinds of heavy physics decouple, and
the vast majority of effects it can produce for the microwave
background can be argued to be smaller than $O(H^2/M^2)$
\cite{InflNoEffect}. Even in this case, however, larger
contributions can be obtained using ordinary inflationary
field-theory examples, where low-energy effects can instead be
suppressed by powers of $m/M$ rather than $H/M$, where $m$ is a
light scale which need not be as small as $H$ \cite{TPI2}.
\end{itemize}

Again the final picture which emerges is encouraging. The criteria
for validity of effective field theories appear to be the same for
gravity as they are in non-gravitational situations. In
particular, for a very broad class of high-energy physics
effective field theory arguments apply, and so theoretical
predictions for the fluctuations in the CMB are robust in the
sense that they are insensitive to most of the details of this
physics. But some kinds of high-energy effects can produce
observable phenomena, and these should be searched for.



\section{Explicit Quantum Calculations}
Although the ideas presented in here have been around for more
than twenty years \cite{physica}, explicit calculations based on
them have only recently been made. Two of these are summarized in
this section.

\subsection{Non-relativistic Point Masses in 3 Spatial Dimensions}
The first example to be considered consists of quantum corrections
to the potential energy, $V(r)$, of gravitational interaction for
two large, slowly-moving point masses separated by a distance $r$.
Working to leading order in source velocities, $v$, we expect the
leading behavior for large source masses to be the Newtonian
gravitational interaction of two classical, static point sources
of energy:
\begin{equation}
 \rho({\bf r}) = \sum_i M_i \, \delta^3({\bf r} - {\bf r}_i) \, .
\end{equation}

Our interest is in the quantum and relativistic corrections to
this Newtonian limit, as described by the gravitational action,
eq.~(\ref{gravaction}), plus the appropriate source action (like,
for instance, eq.~(\ref{NRLagr})). For point sources which are
separated by a large distance $r$ we expect these corrections to
be weak, and so should be calculable in perturbation theory about
flat space. The strength of the gravitational interaction at large
separation is controlled by two small dimensionless quantities,
which suggest themselves on dimensional grounds. Temporarily
re-instating factors of $\hbar$ and $c$, these small parameters
are: $G \hbar/r^2 c^3$ and $G M_i/r c^2$. Both tend to zero for
large $r$, and as we shall see, the first controls the size of
quantum corrections and the second controls the size of
relativistic corrections.\footnote{The point of the
non-relativistic power-counting of the previous section is to show
that the third, large, $r$-independent dimensionless quantity $G
M_i M_j/\hbar c$ does \emph{not} appear in the interaction
energy.}

\subsubsection{Definition of the Potential}
Because there is some freedom of choice in the definition of an
interaction potential in a relativistic field theory, we first
pause to consider some of the definitions which have been
considered. Although more sophisticated possibilities are possible
\cite{OtherPot}, for systems near the flat-space limit a natural
definition of the interaction potential between slowly-moving
point masses can be made in terms of their scattering amplitudes.

Consider, then, two particles which scatter non-relativistically,
with each undergoing a momentum transfer, $\Delta {\bf p}_1 = -
\Delta {\bf p}_2 \equiv {\bf q}$, in the center-of-mass frame. The
most direct definition of the interaction potential, $V({\bf r})$,
of these two particles is to define its matrix elements within
single-particle states to reproduce the full field-theoretical
amplitude for this scattering. For instance, if the
field-theoretic scattering matrix takes the form $\langle f | T |
i \rangle = (2\pi)^4 \, \delta^4(p_f - p_i) \, {\cal A}(q)$, the
potential $V$ would be defined by
\begin{equation}
 \langle f | T | i \rangle = 2 \pi \,
 \delta(E_f - E_i) \,
 \langle f |
 \tilde{V}({\bf q}) | i \rangle \,.
\end{equation}
The position-space potential is then given by $V({\bf r}) = N
(2\pi)^{-3} \int d^3q \, \tilde{V}({\bf q})$. The overall
normalization, $N$, depends on the conventions used for the
normalization of the initial and final states, and is chosen to
ensure the proper form for the Newtonian interaction.

\begin{figure}[t]
  \def\epsfsize#1#2{0.8#1}
  \centerline{\epsfbox{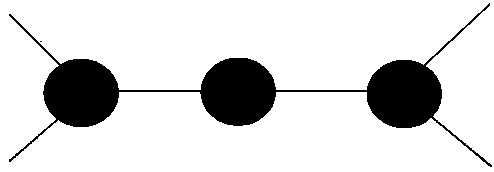}}
  \caption{\it The 1-particle-reducible Feynman graphs relevant to
  the definition of the interaction potential. The blobs represent
  self-energy and vertex corrections.}
  \label{figure:1PRgraphs}
\end{figure}

Several other definitions for the interaction potential have also
been considered by various workers, some of which we now briefly
list.

\medskip\noindent
{\it One-Particle-Reducible Amplitude:}

\medskip\noindent
An alternative definition, followed in refs.~\cite{DonoghueGR} and
\cite{ABS}, is to define the interaction potential in terms of the
one-particle-reducible part of the amplitude, ${\cal A}_{1PR}$,
--- see fig.~(\ref{figure:1PRgraphs})
--- as is commonly done for Quantum Electrodynamics and Quantum
Chromodynamics. The logic of this choice is that because the
graviton propagator varies as $1/q^2$, for small $q^2$ graviton
exchange dominates very-long-distance interactions. It has the
disadvantage that the one-particle-reducible graphs are not
observable in themselves, and so need not form a gauge-invariant
subset. Nevertheless, the results obtained from this definition
can be interpreted \cite{DonoghueGRQED}, \cite{Donoghue-BB1} as
giving the leading quantum corrections to the Schwarzschild,
Kerr-Newman and Reisner-Nordstr\"om metrics.

\medskip\noindent{\it Vacuum Polarization:}

\medskip\noindent
Some early workers defined the interaction potential in terms of
the purely vacuum polarization subset of the 1-particle-reducible
graphs \cite{Duff}. The motivation for such a choice is that these
are the only graphs which would arise for a purely classical
source, which macroscopic objects like planets or stars were
expected to be. It is important to recognize that the
power-counting arguments given earlier necessarily require the
inclusion of vertex corrections at the same order in small
quantities as the vacuum polarization graphs. The necessity for so
doing shows that there is no limit in which a source for the
gravitational field can be considered to be precisely classical.
This non-classicality arises because the gravitational field
itself carries energy, and its quantum fluctuations do not
decouple in the large-mass limit due to the growth which the
gravitational coupling experiences in this limit.

\subsubsection{Calculation of the Interaction Potential}
We now describe the results of recent explicit calculations of the
gravitational potential just defined. A number of these
calculations have now been performed \cite{OtherGR},
\cite{Vambig}, and it is the results of refs.~\cite{DonoghueGR},
\cite{Donoghue-BB1} and \cite{Donoghue-BB2} which are summarized
here.

For any of these potentials, scattering at large distances ($r \to
\infty$) --- {\it i.e.} large impact parameters --- corresponds to
small momentum transfers, ${\bf q}^2 \to 0$. Because corrections
to the Newtonian limit involve the interchange of massless
gravitons, in general scattering amplitudes are not analytic in
this limit. In particular, in the present instance the small-${\bf
q}^2$ limit to the scattering amplitude turns out to behaves as
\begin{equation} \label{qform}
 {\cal A}({\bf q}) = \frac{k_2}{{\bf q}^2} + \frac{k_1}{\sqrt{{\bf
 q}^2}} + k_0 \log \left( {\bf q}^2 \right) + {\cal A}_{\rm
 an}({\bf q}^2) \,
\end{equation}
where ${\cal A}_{\rm an} = A_0 + A_2 {\bf q}^2 + \cdots$ is an
analytic function of ${\bf q}^2$ near ${\bf q}^2 = 0$.

In position space the first three terms of eq.~(\ref{qform})
correspond to terms which fall off with $r$ like $k_2/r$,
$k_1/r^2$ and $k_0/r^3$, respectively. By contrast, the powers of
${\bf q}^2$ in ${\cal A}_{\rm an}$ only contribute terms to
$V({\bf r})$ which are local, inasmuch as they are proportional to
$\delta^3({\bf r})$ or its derivatives. Since our interest is only
in the long-distance interaction, the analytic contributions of
${\cal A}_{\rm an}$ may be completely ignored in what follows.

The power-counting analysis described in earlier sections suggest
that the leading corrections to the Newtonian result come either
from: ($i$) relativistic contributions coming from tree-level
calculations within General Relativity; ($ii$) one-loop
corrections to the classical potential, again using only General
Relativity, or ($iii$) from tree-level contributions containing
precisely one vertex from the curvature-squared terms of the
effective theory, eq.~(\ref{gravaction}). The interaction
potential therefore has the form
\begin{equation}
 V(r) = V_{{\scriptscriptstyle GR},cl}(M_p;r) +
 V_{{\scriptscriptstyle GR},q}(M_p;r) + V_{\rm cs}(a,b,M_p;r) \,,
\end{equation}
respectively corresponding to the contributions of classical
General Relativity, the one-loop corrections within General
Relativity and the classical curvature-squared contributions. The
dependence of this latter term on the quantities $M_p$, $a$ and
$b$ are written explicitly to emphasize which contributions depend
on which parameters. We now describe the result which is obtained
for each of these three types of contribution.


\medskip\noindent {\it Curvature-Squared Terms}

\medskip\noindent
The simplest contribution to dispose of is that due to the
curvature-squared terms.\footnote{Notice that the
curvature-squared terms can no longer be eliminated by performing
field redefinitions once classical sources are included. Instead
they can only be converted into the direct source-source
interactions in which we are interested.} Because these terms are
polynomials in momenta, they contribute only to the analytic part,
${\cal A}_{\rm an}$ of the scattering amplitudes, and so give only
local contributions to the interaction potential which involve
$\delta^3({\bf r})$ or its derivatives. Their precise form is
computed in ref.~\cite{DonoghueGR}, who find
\begin{equation}
\label{GRcorr2}
 V_{\rm cs}(r) = G M_1 M_2 \, B \, \delta^3({\bf r}),
\end{equation}
with $B$ given in terms of the constants $a$ and $b$ of
eq.~(\ref{gravaction}) by $B = 128 \pi^2 G (a + b)$. Since they
contribute only to ${\cal A}_{\rm an}$, we see that these
contributions are necessarily irrelevant to the large-distance
interaction potential.

It is instructive to think of this delta-function contribution due
to curvature-squared terms in another way. To this end, consider
the toy model of a massless scalar field coupled to a classical
$\delta$-function source, whose lagrangian is
\begin{equation} \label{toyHiD}
 - \, {{\cal L} } = \frac{1}{2} (\partial \phi)^2
 + \frac{\kappa}{2} \, (\Box \phi)^2 \,.
\end{equation}
The higher-derivative term proportional to $\kappa$ in this model
is the analogue of the curvature-squared gravitational
interactions. The propagator for this theory satisfies the
equation $(\Box - \kappa \, \Box^2) G_\kappa(x,y) =
\delta^4(x-y)$, which becomes (to linear order in $\kappa$):
$G_\kappa(x,y) \approx G_0(x,y) + \kappa \Box G_0(x,y) = G_0(x,y)
+ \kappa \, \delta^4(x-y)$, where $G_0(x,y)$ is the usual
propagator when $\kappa = 0$. We see the expected result that the
leading contribution to $V(r)$ is purely local in position space
(as might be expected for the low-energy implications of
very-high-energy/very-short-range physics).

This way of thinking of things is useful because it illustrates an
important conceptual issue for effective field theories. Normally
one considers higher-derivative theories to be anathema since
higher-derivative field equations generically have unstable
runaway solutions, and the above calculation shows why these do
not pose problems for the effective field theory. To see why this
is so it is useful to pause to review how the runaway solutions
arise.

At the classical level, runaway modes are possible because of the
additional initial data which higher-derivative equations require.
The reason for their origin in the quantum theory is also easily
seen using the toy theory defined by eq.~(\ref{toyHiD}), for which
at face value the momentum-space scalar propagator would be:
\begin{equation} \label{toyprop}
 -iG(p) \propto \frac{1}{p^2 + \kappa \, p^4} = \frac{1}{p^2} -
 \frac{1}{p^2 + \kappa^{-1}} \,.
\end{equation}
This shows how the higher-derivative term introduces a new pole
into the propagator at $p^2 = - \kappa^{-1}$, but with a residue
whose sign is unphysical (corresponding to a ghost mode with
negative kinetic energy).

The reason these do not pose a problem for effective field
theories is that all of the higher-derivative terms are required
to be treated \emph{perturbatively}, since these interactions are
defined by reproducing the results of the underlying physics
order-by-order in powers of inverse heavy masses, $1/m$. In the
effective theory of eq.~(\ref{toyHiD}) the propagator
eq.~(\ref{toyprop}) must be read as
\begin{equation}
 -iG(p) \propto \frac{1}{p^2} \left( 1  - \kappa \, p^2 + \cdots
 \right) \,,
\end{equation}
since the lagrangian itself is only accurate to leading order in
$\kappa$. The ghost pole does not arise perturbatively in $\kappa
\, p^2$, since its location is up at high energies, $p^2 = -
\kappa^{-1}$. Ref.~\cite{Simon} makes this general argument
explicit for the specific case of higher-derivative gravity
linearized about flat space.

\medskip\noindent{\it Classical General Relativity}

\medskip\noindent The leading contributions for large $r$ due to the
relativistic corrections of General Relativity have the large-$r$
form (with factors of $c$ restored)
\begin{equation}
\label{GRcorr1}
 V_{{\scriptscriptstyle GR},cl}(r) = - \; {G M_1 M_2
 \over r} \left[ 1 + \lambda {G(M_1 + M_2) \over r c^2} + \cdots
 \right]  ,
\end{equation}
where $G = 1/(8 \pi M_p^2)$ is Newton's constant, $M_1$ and $M_2$
are the masses whose potential energy is of interest, and which
are separated by the distance $r$.

The square brackets, $\Bigl[1 + \cdots \Bigr]$, in this expression
represent the relativistic corrections to the Newtonian potential
which already arise within classical General Relativity, and
$\lambda$ is a known constant whose value depends on the precise
coordinate conditions used in the calculation. For example, using
the potential defined by the 1-particle-reducible scattering
amplitude gives \cite{DonoghueGR}, \cite{Donoghue-BB1}
$\lambda_{1PR} = -1$, corresponding to the classical result for
the metric in harmonic gauge, for which the Schwarzschild metric
takes the form
\begin{equation}
 g_{00} = - \frac{(1 - GM/r)}{(1 + GM/r)} = -1 + 2 \left( \frac{GM}{r}
 \right) - 2
 \left( \frac{GM}{r} \right)^2 + \cdots \, .
\end{equation}
Alternatively, using the potential defined by the full scattering
amplitude, ${\cal A}_{\rm tot}$, instead gives \cite{Donoghue-BB2}
$\lambda_{\rm tot} = + 3$. It is natural that different values for
$\lambda$ are obtained when different definitions for $V$ are
used, since these different definitions contribute differently to
physical observables (on which all calculations must agree).

There is another ambiguity in the definition of the potential
\cite{Vambig}, which is related to the freedom to redefine the
coordinate $r$, according to $r \to r' = r \left[1 + a G M/r +
\cdots \right]$. Of course, such a coordinate change should drop
out of physical observables, but how this happens in this case
involves a subtlety. The main point is that the low-energy
effective lagrangian for the non-relativistic particles contains
\emph{two} terms of the same size at subleading order in the
relativistic expansion, having the schematic form: $\Delta {\cal
L} = \lambda (G M^2/r)(GM/r) + \lambda' (G M/r) (M v^2)$, where
$M$ and $v$ are the mass and velocity of the non-relativistic
particle of interest. The main point is that the constants
$\lambda$ and $\lambda'$ are redundant interactions in the sense
defined earlier, inasmuch as all physical observables only depend
on a single combination of these two constants. Observables only
depend on one combination because the other combination can be
removed by performing the coordinate transformation $r \to r [1 +
a GM/r + \cdots]$, as above. From this we see that the coefficient
$\lambda$ of $GM/r^2$ obtained for $V(r)$ can also differ from one
another, provided that the coefficient $\lambda'$ also differs in
such a way as to give the same results for physical observables.

\medskip\noindent {\it One-Loop General Relativity}

\medskip\noindent
The final term in $V(r)$ arises from the one-loop contribution as
computed within General Relativity, which is extracted by
calculating the one-loop corrections to the scattering amplitude,
${\cal A}_q$. Although these corrections typically diverge in the
ultraviolet, on general grounds such divergences contribute only
polynomials in momenta, and so can contribute only to the
non-relativistic amplitude's analytic part, ${\cal A}_{\rm
an}({\bf q}^2)$. Indeed, this is required for the one-loop
divergences to be absorbed by renormalizing the effective
couplings $a$ and $b$ of the higher-curvature terms of the
gravitational action, eq.~(\ref{gravaction}). (The necessity for
renormalizing $a$ and $b$ in addition to Newton's constant at one
loop reflects the fact that General Relativity is not
renormalizable. Still higher-curvature terms would be required to
absorb the divergences at two loops and beyond.)

It follows from this observation that to the extent that we focus
on the long-distance interactions in $V(r)$, to the order we are
working these must be ultraviolet finite since they receive no
contribution from the amplitude's analytic part. This means the
leading quantum implications for $V(r)$ are unambiguous
predictions which are not complicated by the renormalization
procedure.

Explicit calculation shows that the non-analytic part of the
quantum corrections to scattering are proportional to $\log {\bf
q}^2$, and so the leading one-loop quantum contribution to the
interaction potential is (again re-instating powers of $\hbar$ and
$c$)
\begin{equation}
\label{GRcorr3}
 V_{{\scriptscriptstyle GR},q}(r) = - \; {G M_1 M_2 \over r} \left[ 1 +
 \xi \, {G \hbar \over r^2 c^3} + \cdots \right],
\end{equation}
where $\xi$ is a calculable number. If the potential is computed
using only the one-particle-reducible scattering amplitude, the
result for pure gravity is \cite{Donoghue-BB1}:
\begin{equation}
 \xi_{1PR} = - \frac{167}{30 \pi}  \,.
\end{equation}
Notice that this corrects an error in the earlier result for the
same quantity, given in ref.~\cite{DonoghueGR}. If, instead, the
full amplitude ${\cal A}_{\rm tot}$ is used to define the
interaction potential, ref.~\cite{Donoghue-BB2} find
\begin{equation}
 \xi_{\rm tot} = + \frac{41}{10 \pi} \, .
\end{equation}
It is argued in ref.~\cite{Donoghue-BB2} that these one-loop
results for $\xi$ do not suffer from ambiguity due to the freedom
to perform redefinitions of the form $r \to r [1 + a G^2/r^2 +
\cdots ]$.

\subsubsection{Implications}
It is remarkable that the quantum corrections to the interaction
potential can be so cleanly identified. In this section we
summarize a few general inferences which follow from their size
and dependence on physical parameters like mass and separation.

Conceptually, the main point is that the quantum effects are
calculable, and in principle can be distinguished from purely
classical corrections. For instance, the quantum contribution
(\ref{GRcorr3}) can be distinguished from the classical
relativistic corrections (\ref{GRcorr1}) because the quantum and
the relativistic terms depend differently on $G$ and the masses
$M_1$ and $M_2$. In particular, relativistic corrections are
controlled by the dimensionless quantity $GM_{\rm tot}/rc^2$,
which is a measure of typical orbital velocities, $v^2/c^2$. The
leading quantum corrections, on the other hand, are
$M$-independent and are controlled by the ratio $\ell_p^2/r^2$,
where $\ell_p = (G\hbar/c^3)^{1/2} \sim 10^{-35}$ m is the Planck
length.

Although the one-particle-reducible contributions need not be
separately gauge-independent, refs.~\cite{Donoghue-BB1} and
\cite{DonoghueGRQED} argue that they may be usefully interpreted
as defining long-distance quantum corrections to the metric
external to various types of point sources. Besides obtaining
corrections to the Schwarzschild metric in this way, they do the
same for the Kerr-Newman and Reissner-Nordstr\"om metrics by
incorporating spin and electric charge into the non-relativistic
quantum source. Because the quantum corrections they find are
source-independent, these authors suggest they be interpreted in
terms of a running Newton's constant, according to
\begin{equation}
 G(r) = G \left[ 1 - \frac{167}{30 \pi} \, \left( \frac{G}{r^2}
 \right) + \cdots \right] \,.
\end{equation}

Numerically, the quantum corrections are so miniscule as to be
unobservable within the solar system for the forseeable future.
Table 1 evaluates their size using for definiteness a solar mass,
$M_\odot$, and with $r$ chosen equal to the solar radius, $R_\odot
\sim 10^9$ m, or the solar Schwarzschild radius, $r_s = 2
GM_\odot/c^2 \sim 10^3$ m. Clearly the quantum-gravitational
correction is numerically extremely small when evaluated for
garden-variety gravitational fields in the solar system, and would
remain so right down to the event horizon even if the sun were a
black hole. At face value it is only for separations comparable to
the Planck length that quantum gravity effects become important.
To the extent that these estimates carry over to quantum effects
right down to the event horizon on curved black-hole geometries
(more about this below) this makes quantum corrections irrelevant
for physics outside of the even horizon, unless the black hole
mass is as small as the Planck mass, $M_{\rm hole} \sim M_p \sim
10^{-5}$ g.

\vspace{3mm}
\begin{center}
\begin{tabular}{lcc}
& $r = R_\odot$ & $r = 2 GM_\odot/c^2$ \\
&&\\
$GM_\odot/rc^2$ & $10^{-6}$ & $0.5$ \\
$G\hbar/r^2c^3$ & $10^{-88}$ & $10^{-76}$
\end{tabular}
\end{center}

\begin{quote}
{\footnotesize {\bf Table 1:} {\sl The generic size of
relativistic and quantum corrections to the Sun's gravitational
field.}}
\end{quote}

Of course, the undetectability of these quantum corrections does
not make them unimportant. Rather, the above calculations
underline the following three conclusions:

\begin{itemize}
\item One need not throw up one's hands when contemplating quantum
gravity effects, because quantum corrections in gravity are often
unambiguous and calculable.
\item Although the small size of the above quantum corrections
in the solar system mean that they are unlikely to be measured,
they also show that the great experimental success of classical
General Relativity in the solar system should also be regarded as
a triumph of {\it quantum} gravity! Classical calculations are not
a poor substitute for some poorly-understood quantum theory, they
are rather an extremely good approximation for which quantum
corrections are exceedingly small.
\item Despite the above two points, the mysteries of quantum gravity
remain real and profound. But the above calculations show that
these are high-energy (or short-distance) mysteries, and so point
to cosmological singularities or primordial black holes as being
the places to look for quantum gravitational effects.
\end{itemize}

\subsection{Co-Dimension Two and Cosmic Strings}
A second explicit calculation of quantum corrections within
General Relativity has been done for the gravitational field of a
cosmic string, which for our purposes is a line distribution of
mass characterized by a mass-per-unit-length, $\rho$. This system
naturally suggests itself as a theoretical laboratory for
computing quantum effects because its classical gravitational
field is extremely simple.

The classical field due to a line distribution of mass is simple
for the following reason. Because of the symmetry of the mass
distribution, the calculation of the gravitational field it
produces is effectively a 2+1 dimensional problem. If the exterior
to the mass distribution is empty, we seek there a solution to the
vacuum Einstein equations $R_{\mu\nu} = 0$. But it is a theorem
that in 2+1 dimensions any geometry which is Ricci flat must also
be Riemann flat: $R_{\mu\nu\lambda\rho} = 0$! Superficially this
appears to lead to the paradoxical conclusion that long, straight
cosmic strings should not gravitate.

This conclusion is not quite correct, however. Although it is true
that the vanishing of the Riemann tensor implies no tidal forces
for test particles which pass by on the \emph{same} side of the
string, test particles are influenced to approach one another if
they pass by on \emph{opposite} sides of the string. The reason
for this may be seen by more closely examining the spacetime's
geometry near the position of the cosmic string. The boundary
conditions at this point require that spacetime there to resemble
the tip of a cone, inasmuch as an infinitely thin cosmic string
introduces a $\delta$-function singularity into the curvature of
spacetime. This implies that the flat geometry outside of the
string behaves globally like a cone, corresponding to the removal
a defect angle, $\Delta\theta = 8\pi G \rho$ radians, from the
external geometry. This conical geometry for the external
spacetime is what causes the focussing of trajectories of pairs of
particles which pass by on either side of the string \cite{Deser}.

The above considerations show that the gravitational interaction
of two cosmic strings furnishes an ideal theoretical laboratory
for studying quantum gravity effects near flat space. Since the
classical gravitational force of one string on the other vanishes
classically, its \emph{leading} contribution arises at the quantum
level. Consider, for instance, the interaction energy
per-unit-length, $u_{\rm int}$, of two straight parallel strings
separated by a distance $a$. This receives no contribution from
the Einstein-Hilbert term of the effective action, for the reasons
just described. Furthermore, just as for point gravitational
sources, higher-curvature interactions only generate contact
interactions, and so are also irrelevant for computing the
strings' interactions at long range. The leading contribution
therefore arises at the quantum level, and must be ultraviolet
finite.

These expectations are borne out by explicit one-loop
calculations, which have been computed \cite{Wise} for the case of
two strings having constant mass-per-unit-length, $\rho_1$ and
$\rho_2$. The result obtained is (again temporarily restoring the
explicit powers of $\hbar$ and $c$):
\begin{equation}
 u_{\rm int}(a) = \left( \frac{24 \hbar}{5 \pi c^3} \right) \frac{ G^2 \rho_1
 \rho_2}{a^2} \, ,
\end{equation}
whose sign corresponds to a repulsive interaction.



\section{Conclusions}
\label{section:suggestions}

The goal of this review has been to summarize the modern picture
of quantum gravity, within which the perturbative
non-renormalizability of General Relativity is recognized as being
a particular instance of a more general phenomena: the widespread
application of non-renormalizable quantum field theories
throughout many branches of physics. Regarding quantum gravity in
this way shows how quantitative predictions can be made: one must
simply apply the rules of effective field theories, which are
known to give an accurate description of experiments in low-energy
nuclear, particle, atomic and condensed matter physics.

Thinking of General Relativity as an effective theory in this way
is not a new development, and underlies most approaches to quantum
gravity either explicitly or implicitly. Neither is it new to
calculate explicitly the behaviour of quantum fields in curved
space (sometimes including the graviton). What \emph{is} new (over
the last few years) is proceeding beyond the qualitative statement
that General Relativity is an effective theory to obtain the
quantitative next-to-leading predictions within a controlled
semiclassical approximation. Although much of the mechanics of
such calculations leans on experience obtained when calculating
with quantum fields in curved space, the crucial new difference is
the quantitative power-counting arguments which identify precisely
which quantum effects contribute to any given order in small
quantities.

What emerges from this summary is a snapshot of a work which is
very much still in progress. The following loom large among the
missing results.
\begin{itemize}
\item
Although a general statement of the power-counting result for very
light, relativistic particles near flat space has been known for
some time \cite{DonoghuePC}, \cite{Ode}, the central general
power-counting results are not yet demonstrated to all orders for
the most interesting case for practical calculations: the
gravitational interactions of very massive, non-relativistic
sources which are weakly interacting gravitationally ({\it i.e.}
in spacetimes which are perturbatively close to Minkowski space).
\item
Although the theoretical tools exist, in the form of the operator
product expansion, similarly general power-counting arguments are
not yet given for quantum fluctuations about more general curved
spaces.
\item
Some explicit calculations which are cast within the
power-counting framework have been done, but many more can surely
be done.
\end{itemize}

There can be little doubt that quantum effects are extremely small
in the classical systems for which gravitational measurements are
possible (like the solar system), but this need not undermine the
motivation for their computation. The point of such calculations
is not their relevance for practical experiments (we wish!).
Rather, their point is conceptual. It is only through the careful
calculation of quantum effects that the theory of their size can
be solidly established. In particular, any precise comparison
between observations and the predictions of classical gravity is
ultimately incomplete unless the quantitative size of the quantum
corrections is explicitly established, as a systematic, all-orders
power-counting argument would do.

Furthermore, we can always hope to get lucky, even if only
theoretically. A clean understanding of how the size of quantum
corrections depends on the variables (mass, size, separation, {\it
etc.})in a given system, one might hope to find
larger-than-generic quantum phenomena in special systems. Even if
these lie beyond the reach of present-day experimenters, they may
furnish instructive theoretical laboratories within which
differing approaches to quantum gravity might be more starkly
compared.

In the last analysis, I hope the reader has become convinced of
the utility of effective field theory techniques, and that the
effective field theory point of view lifts the experimental
triumphs of classical General Relativity to precision tests of the
leading-order implications of the Quantum Theory of Gravity.



\section{Acknowledgements}
\label{section:acknowledgements}

I wish to thank the editors of \emph{Living Reviews} for the
invitation to write this article, and for their great patience in
waiting for its delivery. I am also grateful to John Donoghue and
Ted Jacobson for their many useful comments on an early draft of
this review. My own ideas on this topic are heavily indebted to
Steven Weinberg, as conveyed through his graduate lectures on
Quantum Field Theory (and now by his textbooks on the subject).

\newpage




\end{document}